%% file: implementations.tex
\documentclass[IEEEtran,conference]{implementations}

\pdfoutput=1

\renewcommand{\prm}[1]{{({}{#1})}}

\title{Decomposing, Comparing, and Synthesizing Access Control Expressiveness Simulations \inlong{\\\smaller[1]\textit{(Extended Version)}}}

\begin{document}

\maketitle

\input{abstract}

\input{introduction}

\input{expressiveness}

\input{implementable}

\input{properties}

\input{positioning}

\input{selecting}

\input{conclusion}

\input{implementations.bbl}

\inlong{
\appendices
\input{queryproperties}

\input{proofs}

}

\end{document}

%% file: abstract.tex
\begin{abstract}
Access control is fundamental to computer security, and has thus been the
subject of extensive formal study.  In particular, \emph{relative expressiveness
analysis} techniques have used formal mappings called \emph{simulations} to
explore whether one access control system is capable of emulating another,
thereby comparing the expressive power of these systems. Unfortunately, the
notions of expressiveness simulation that have been explored vary widely, which
makes it difficult to compare results in the literature, and even leads to
apparent contradictions between results. Furthermore, some notions of
expressiveness simulation make use of non-determinism, and thus cannot be used
to define mappings between access control systems that are useful in practical
scenarios. In this work, we define the minimum set of properties for an
\emph{implementable} access control simulation; i.e., a deterministic ``recipe''
for using one system in place of another. We then define a wide range of
properties spread across several dimensions that can be enforced on top of this
minimum definition. These properties define a taxonomy that can be used to
separate and compare existing notions of access control simulation, many of
which were previously incomparable. We position existing notions of simulation
within our properties lattice by formally proving each simulation's equivalence
to a corresponding set of properties.  Lastly, we take steps towards bridging
the gap between theory and practice by exploring the systems implications of
points within our properties lattice.  This shows that relative expressive
analysis is more than just a theoretical tool, and can also guide the choice of
the most suitable access control system for a specific application or scenario.
\end{abstract}

%% file: introduction.tex
\section{Introduction} \label{sec:introduction}

Access control is foundational to computer security and, as such, has been the
topic of extensive formal study. Much of this work has focused on comparing
different techniques for representing and enforcing access control, deemed
access control \emph{models}, \emph{systems}, or \emph{schemes}. By far the most
common type of comparative study in access control techniques is the
expressiveness simulation
(e.g.,~\cite{ALS92,Sandhu92,SG93,SG94,Ganta96,ALS96,SM98,MS99,OSM00,CDM01,BCF03,TL04,TL07,HMG13}).
A simulation is a formal mapping from, say, system \system{S} to system
\system{T} that proves \system{T} is at least as expressive as \system{S}: that
is, \system{T} possesses the raw capability to be used in operating environments
in place of \system{S}.

However, the formal definitions of the various simulations used in the
literature vary widely. Different simulations have been used to prove various
types of results, ranging from very specific properties about whole ranges of
models (e.g., monotonic access control models with multi-parent creation cannot
be simulated by monotonic models with only single-parent creation~\cite{ALS96})
to the ability to replace certain specific models with others in practice (e.g.,
role-based access control can be configured to enforce mandatory and
discretionary policies~\cite{OSM00}). However, this disparity in the goals of
these works has led to many different definitions of access control simulation,
often tailored to the particular result sought. It has been shown that these
different simulations prove wildly different notions of expressiveness, often
not preserving any particular security properties~\cite{TL07}.

Furthermore, not all of these notions of simulation are practically useful. For
instance, some make use of non-determinism, manipulating the policy differently
depending on what future queries will be asked. While this may allow a theorist
to show that system \system{T} is capable of doing all the things \system{S} is,
if a practitioner wants to use system \system{T} in place of system \system{S},
she needs a deterministic procedure for doing so.

In this work, we build a taxonomy for expressiveness simulations based on the
simulation properties that they satisfy. We determine the minimum requirements
for a mapping to be \emph{implementable}, or applicable toward using one
system in place of another in practice. We use these requirements to construct a
general definition of implementable simulation, and provide a taxonomy of
additional restrictions on this definition for simulations that enforce more
stringent properties. We then position existing simulations from the literature
within this lattice, providing the first such comparison in the literature.

To this end, we make the following contributions.
\begin{parlist}

\item{Definition of implementable access control simulation} We propose a
general definition of an \emph{implementable} access control mapping that is
broad enough to encompass much of the wide range of existing access control
simulations, yet precise enough to guarantee implementability.  Intuitively, an
\emph{implementable} simulation of \system{S} in \system{T} shows that
\system{T} can accomplish everything \system{S} can, and deterministically shows
\emph{how} (\Cref{sec:implementable}).

\item{Lattice of simulation properties} We decompose and expand upon the
properties enforced by various access control simulations from the literature,
forming a lattice relating the range of access control simulations to one
another. This lattice allows us to formally compare the guarantees offered by
existing notions of access control simulation (many of which were not formerly
known to be comparable) and points to unexplored combinations of properties that
can yield different expressiveness results (\Cref{sec:properties}).

\item{Positioning of existing simulations} We construct formal proofs
positioning existing notions of access control simulation within our lattice of
simulation properties, including a comparative discussion of simulations that
previously seemed incomparable. We thus systematize the formal relationships
between previously-published simulations, allowing reconciliation of previously
disparate expressiveness knowledge (\Cref{sec:positioning}).

\item{Selecting simulation properties} We observe that many of the dimensions
upon which our simulation property lattice is built have implications for the
use of simulations for satisfying real-world requirements using existing access
control systems (e.g., required storage, whether data structures must be locked
for concurrent usage). Thus, in addition to positioning existing notions of
simulation within our lattice of properties, we assist in creating new notions
of simulation by selecting the properties that should be enforced in an
expressiveness analysis based upon the scenario in which an eventual access
control deployment will occur. To this end, we discuss in detail various
interactions between simulation properties, the results of enforcing different
properties, and how a specific deployment scenario dictates which properties are
relevant (\Cref{sec:selecting}).

\end{parlist}

We begin by providing background on the goals of and techniques used in relative
expressiveness analysis.

%% file: expressiveness.tex
\section{Relative Expressiveness Analysis} \label{sec:expressiveness}

In this \lcnamecref{sec:expressiveness}, we describe how relative expressiveness
analysis is conducted, survey the history of the technique, and point out the
wide variety in existing access control expressiveness simulations.

\subsection{Motivating Examples}
\label{sec:expressiveness:motivation}

An access control system's \emph{expressiveness} (or expressive power) is a
measure of the range of policies that it can represent and the transformations
it can make to those policies. Statements of \emph{relative} expressiveness
state that one system is capable of replacing another (that is, it can represent
all the same policies and transform them in equivalent ways). Assume, for
instance, that an organization is considering transitioning from one access
control solution to another, in order to accommodate evolving requirements. The
organization may have specific desired features for this new access control
system, but it certainly must be able to represent all of the policies that the
existing system can, or it would not be a suitable replacement. Thus, this
organization is searching for a new system that is \emph{at least as expressive
as} its old system.

Another use of relative expressiveness is in suitability analysis. Prior work
has noted that \emph{practically} evaluating an access control system must take
into account the application in which the system is to be used, as well as
additional cost metrics (e.g., computation, ease of use). This analysis problem
has been identified as a system's \emph{suitability} to a particular
application~\cite{codaspy,sacmat}. Suitability analysis formalizes an
application's access control requirements (a \emph{workload}), and uses
expressiveness to prove that an access control system can satisfy those
requirements. Assume, in this case, that the aforementioned organization is
choosing an initial access control system for a new collection of data.
Comparing the candidates' relative expressiveness is not particularly
enlightening, since the most expressive system may not be the most suitable; the
organization should instead formalize their access control workload and use
relative expressiveness analysis to identify which of the candidates are
\emph{expressive enough} to satisfy this workload. Thus, while work in
suitability analysis has shown that expressive power alone is insufficient for
evaluating an access control system, expressiveness is a fundamentally important
component of a more general suitability analysis workflow: one cannot determine
which access control system is \emph{best} for a particular use case without
first determining which are \emph{capable of satisfying} that use case.

\subsection{Prior Work}

Relative expressiveness analysis generally starts by formalizing a pair of
access control systems as state machines. These state machines include, at a
minimum: a set of states, each of which encapsulates a snapshot of the access
control system's data structures; a procedure describing how to interpret the
states' data structures to determine which authorization requests are granted;
and a set of commands, used to manipulate the data structures and thus
transition between states. Some formalisms for access control systems also
include additional queries beyond access requests~\cite{TL07,HMG13}. A
\emph{simulation}, then, is a structure that proves \system{T} is at least as
expressive as \system{S}---or, that \system{T} can be used in place of
\system{S}. The term \emph{simulation} is rather vague, here, and for good
reason: various notions of simulation in the literature have meant very
different things (e.g., What type of behavior must be simulated? How closely
must \system{T} represent the information in \system{S}?), and as a result have
implied very different types of expressiveness results.

The works of Sandhu, Ganta, Munawer, and
Osborn~\cite{Sandhu92,SG93,SG94,SM98,MS99,OSM00} include some of the earliest
access control simulations. In these works, a simulation of \system{S} in
\system{T} must show that a permission can be granted in \system{S} if and only
if it can also be granted in \system{T}. No other formal properties are
enforced, though in some cases additional properties become part of the \emph{de
facto} definition of simulation. For instance, while there is no requirement for
\system{T} to have a state equivalent to each \system{S} state (merely for
\system{T} to be able to grant each access that \system{S} does, in some state),
the example simulations all include methods for mapping each \system{S} state to
a \system{T} state (as this is the simplest way to show the required property).
In addition, although the definition does not prohibit the use of an unbounded
number of \system{T} commands to simulate a single \system{S} command, Sandhu
and Munawer~\cite{SM98} only use simulations in which an \system{S} command is
simulated using a constant number of \system{T} commands.

Ganta's PhD dissertation~\cite{Ganta96} attempts to formalize a more rigorous
notion of expressiveness simulation. In his simulation, the state correspondence
is explicit, requiring that each state in \system{S} have a corresponding state
in \system{T} that grants all the same accesses (at least, all those that exist
in \system{S}---those that exist in \system{T} but not in \system{S} are
unconstrained). In addition, to ensure that \system{T} cannot grant accesses
that \system{S} cannot, any state that can be entered in \system{T} must also
have a corresponding reachable state in \system{S}. Finally, to ensure accesses
in \system{T} cannot be combined in ways that cannot occur in \system{S}, the
following restriction is made: when simulating a \system{T} command in
\system{S}, multiple commands may be used, but each state along the way must
allow \emph{either} a subset of the accesses of the start state or a subset of
the accesses of the end state. Thus, no two accesses can be allowed in the same
state in \system{T} that are not allowed in a single state in \system{S}.

Ammann, Lipton, and Sandhu~\cite{ALS92,ALS96} took a different (and much more
strict) approach to more rigorously defining a simulation. First, they describe
a strict state correspondence that requires \system{T} to represent its states
with the same sets and relations as \system{S}, and for these sets to have
identical contents in corresponding \system{T} and \system{S} states. In other
words, \system{T} cannot include additional elements in any sets that \system{S}
uses (although additional, distinct sets may be stored). For example, one could
simulate the state \(\{U = \set{a, b}, V = \set{c}\}\) with state \(\{U =
\set{a, b}, V = \set{c}, W = \set{\tup{a, d}, \tup{b, d}}\}\), but not with
\(\{U = \set{a, b}, V = \set{c, d}\}\). Given this notion of state
correspondence, a simulation then shows that \system{T} can reach a state
corresponding to each reachable \system{S} state, and cannot reach any state
that does not have a reachable corresponding state in \system{S}. This strict
notion of simulation is used to show that monotonic, multi-parent systems are
more expressive than monotonic, single-parent systems (e.g., there are monotonic
multi-parent systems that cannot be simulated by any monotonic single-parent
system).

Chander, Dean, and Mitchell~\cite{CDM01} restrict the definition of simulation
in a different way. Rather than force a more strict state correspondence (the
static portion of the simulation), they more tightly restrict the way the
simulation handles the system as it executes (i.e., the command mapping). In
these simulations, the state correspondence is comparatively lax: to simulate an
\system{S} state, a \system{T} state must allow and deny all the same
authorization requests as its corresponding \system{S} state. Additional
requests can exist in \system{T} and are unconstrained, but all requests
corresponding to those in \system{S} must have the same value in corresponding
states. However, the process for simulating an \system{S} command using
\system{T} commands must be independent of the state: it cannot execute a
\system{T} command for each user, or otherwise inspect the state when
determining what commands should be executed. In addition, in the strong form of
simulation, each \system{S} command must be simulated with a single \system{T}
command. They then compare the expressiveness of access control lists, trust
management, and two forms of capability systems (all systems studied in forms
with and without revocation and delegation).

Tripunitara and Li~\cite{TL04,TL07} noted that the existing notions of
simulation did not correspond directly to any particular safety analysis
questions, and thus a simulation of any of these types does not make any
particular safety guarantees. They formalize \emph{compositional security
analysis} (intuitively, determining whether a certain set of access control
queries will always, never, or sometimes become true in any reachable state),
which is a generalization of simple safety analysis~\cite{hru}. They then
present a notion of simulation tailor-made to preserve these types of analysis
questions.

Their simulation, called the state-matching reduction, considers a broader range
of queries than only authorization requests, placing the strictness of its state
correspondence somewhere between the work of Ammann, Lipton, and Sandhu and that
of Chander, Dean, and Mitchell. The state-matching reduction maps each query
\(\query^\system{S}\) in \system{S} to a single query \(\query^\system{T}\) in
\system{T}, and the simulation must determine the value of \(\query^\system{S}\)
in any state in \system{T} by checking the value of \(\query^\system{T}\).
Finally, reachability constraints ensure that \system{T} can reach a state
corresponding to each reachable \system{S} state, and cannot reach any state
that does not have a reachable corresponding state in \system{S}. Tripunitara
and Li prove that this notion of simulation preserves compositional security
analysis instances: that is, if there exists a state-matching reduction from
\system{S} to \system{T}, then any compositional security analysis instance has
the same truth value in both systems. Tripunitara and Li's reductions have since
been used to analyze role-based access control~\cite{triplirolebased} and prove
that newly-proposed systems are more expressive than certain existing
systems~\cite{tba}.

Work by Hinrichs et al.~\cite{HMG13} recognizes the value of the state-matching
reduction but claims that, in practice, not all scenarios require the
preservation of all possible compositional security analysis instances (nor are
these the only types of safety properties that are ever relevant). They present
\emph{parameterized expressiveness}, which defines a baseline set of simulation
properties, and provides several additional properties that can be enforced atop
the baseline to provide additional guarantees. The base simulation uses the same
query-based state correspondence as Tripunitara and Li, but relaxes the query
mapping to allow it to consult multiple \system{T} queries to determine the
value of an \system{S} query during simulation. Further properties enforced above
this baseline include using the identity query mapping for authorization
requests (to ensure that \system{T}'s authorization questions are the queries
being used to simulate \system{S}'s authorization requests), forbidding string
manipulations (to prohibit the state mapping from using arbitrary encodings to
store information in the contents of strings such as user names), and
restricting the command mapping from mapping non-administrative commands in
\system{S} to administrative commands in \system{T}. This framework has since
been used to evaluate the suitability of certain general-purpose access control
systems for various unique, application-specific
requirements~\cite{codaspy,sacmat}.

\subsection{Usage and Implications}

Unfortunately, there are several indications that research on expressiveness
analysis is being held back by the inability to reconcile the vastly different
notions of expressiveness simulations and the disconnect between the properties
preserved by a simulation and those that are important to a practical
deployment. Several works have demonstrated scenarios in which static notions of
expressiveness indicate two systems are equally capable of satisfying a set of
operational requirements, but in practice they are better-suited to very
different deployment scenarios~\cite{naldurg03,codaspy}. Bourdier et al.\ point
out the existence of several competing techniques for expressiveness analysis,
none of which consider the deployment. They approach one facet of this problem
by proposing a formalism for access control systems that can more easily be
transformed into implementations using rewrite-based tools~\cite{bourdier12}.
Several others simply express a desire to use expressiveness analysis, but never
do so, presumably due to the complexities of selecting and using the right
notion of simulation~\cite{payne07,gsis}.

A group at the National Institute of Standards and Technology has developed
Policy Machine, an attempt at a universal access control system (one that can
represent any policy via only configuration changes)~\cite{policymachine}.
However, in evaluating Policy Machine's success, they avoid formally proving its
expressiveness and instead show informal mappings that demonstrate how one might
use Policy Machine to represent several existing access control systems'
policies~\cite{pmcompose}. Soon after, the group published a report bemoaning
the lack of quality metrics for evaluating access control systems, noting that,
in access control, ``one size does not fit all,'' and thus said metrics must
consider the deployment scenario~\cite{nist}.

This overview illustrates that while each notion of expressiveness simulation
has been used to prove various results, the body of knowledge is troublesome to
interpret and utilize due to the wide variation in the properties required by
each simulation. In this work, we fill this void in the literature by (1)
proposing a minimal definition of simulation that satisfies properties
guaranteeing that its results are practically useful; (2) presenting a set of
additional properties that more strict simulations can enforce; and (3)
categorizing the above notions of simulation based on the properties that they
enforce. We make the additional contribution of (4) discussing relationships
between attributes of a deployment scenario and the practical effects of
enforcing simulation properties, thus assisting analysts in selecting the most
relevant properties (and therefore conducting the most relevant form of
expressiveness analysis) for the environment in which an access control system
will be deployed.

%% file: implementable.tex
\section{Implementable Expressiveness Simulations} \label{sec:implementable}

In this \lcnamecref{sec:implementable}, we give requirements for a simulation to
be \emph{implementable} and define our general formulation of relative
expressiveness analysis through the lens of implementability.

\subsection{Implementability Requirements}
\label{sec:implementable:requirements}

In this work, we aim to consider expressiveness simulations that are
\emph{implementable}: i.e., practically useful for making decisions about which
system is most suitable for a particular deployment. Implementability enforces
the following intuition: if a system \system{T} is at least as expressive as
\system{S}, then one should be able to determine a general way to use \system{T}
in place of \system{S}. Thus, we define a minimal set of properties for an
expressiveness mapping to be considered implementable.

\begin{parlist}

\item{State mapping}
In order to use \system{T} in place of \system{S}, it must be possible to
(uniquely) determine which \system{T} state to use in place of a particular
\system{S} state. Thus, the state mapping must be a function from the simulated
system states to the simulating system states.%
\footnote{It is possible that multiple states in \system{S} can be represented
using the same state in \system{T}. Thus, we do not require the state mapping to
be an injection. Furthermore, there may be states in \system{T} that are not
used to simulate \system{S}, and thus the state mapping need not be an
surjection.}

\item{Command mapping} To use \system{T} in place of \system{S}, it must be
possible to execute commands in \system{T} that are equivalent to the commands
in \system{S}. It is not necessarily the case that each \system{S} command can
be simulated using a single \system{T} command, so we require a function from
\system{S} commands to \emph{sequences of} \system{T} commands.%
\footnote{Not \emph{sets} of \system{T} commands, as commands may appear
multiple times; and not \emph{bags} of \system{T} commands, as order matters.}
Finally, it may be necessary to map an \system{S} command differently depending
upon the state in which it is intended to be executed. Since using \system{T} in
place of \system{S} means we only have a \system{T} state to inspect during
execution, this function should map an \system{S} command and a \system{T} state
to a sequence of \system{T} commands.

\item{Query decider} For some simulations of \system{S} in \system{T}, we may
only care that \system{T} allows the same set of accesses that \system{S} would.
However some types of simulations may allow the overriding of \system{T}'s
default method of deciding granted permissions (e.g., adding the additional
requirement that the requesting user is a member of the \textstt{REAL\_USERS}
group, to distinguish from other data stored in the user-set). While some types
of simulations do not allow this, to remain general we simply require a function
that maps each \system{S} query and \system{T} state to either true or false. In
some formalisms, this only includes the queries requesting access, while in
other cases other types of queries are allowed (e.g., ``Is user \(u\) a member
of role \(r\)?'').

\end{parlist}

We use these requirements to motivate our definition of the general case of
implementable relative expressiveness.

\subsection{Expressiveness Mappings}
\label{sec:implementable:definition}

To define relative expressiveness mappings, we must first define the state
machines that represent access control systems. Since we aim to compare existing
expressiveness simulations, we use a formalism for these structures that remains
similar to existing work, e.g.,~\cite{CDM01,TL04,TL07}.

An access control system is formalized as a state machine belonging to a
particular access control \emph{model}. An access control model formalizes the
way in which the access control system will store and interpret information to
make access control decisions. Its data structures are formalized as a set of
access control \emph{states}, and its methods for determining whether to allow
or deny inquiries as a set of \emph{authorization requests}. The value of all
requests in a state (whether they are allowed or denied) defines the access
control policy, or \emph{theory}, to be enforced in that state.
\begin{definition} \label{def:model}
An \emph{access control model} is defined as \(\model = \tup{\states,
\requests}\), where \states is the set of states and \requests is the set of
authorization requests, where each request \(\request \in \requests\) is a
function \(\states \to \set{\true, \false}\). The entailment (\(\entails\)) of a
request is defined as \(\state \entails \request \defas \request\prm{\state} =
\true\).
\end{definition}

For example, consider a simple role-based access control model whose states are
defined over sets \(U\) of users, \(P\) of permissions, and \(R\) of roles, as
well as the user assignment \(UR \subseteq U \times R\) and permission
assignment \(PA \subseteq R \times P\). The requests in this model are of the
form ``Is \(u\) authorized for \(p\)?,'' which is \true if \(\exists r: \tup{u,
r} \in UR \land \tup{r, p} \in PA\).

When we refer to the \emph{size} of a state, we are referring to the size of its
decomposition into primitive objects (e.g., users and roles) and tuples (e.g.,
entries in a user assignment relation).
\begin{definition}
Given an access control model \(\model = \tup{\states, \requests}\) and a state
\(\state \in \states\), the \emph{set decomposition} of \state is denoted
\(\statedecompose{\state}\), and refers to the ``set of sets'' forming \state,
in which \state is represented as being comprised of primitive sets and
relations.
\end{definition}
Thus, the size of an access control state \state is defined as \(\card{\state} =
\sum_{\statesets\in\statedecompose{\state}}\card{\statesets}\). For example, if
\(\statedecompose{\state} = \{U = \set{u_1}, R = \set{r_1, r_2}, UR =
\set{\tup{u_1, r_1}, \tup{u_1, r_2}}\}\), then \(\card{\state} = \card{U} +
\card{R} + \card{UR} = 5\).

An access control \emph{system} expands on a model by providing methods of
transforming the current state and additional methods of querying the states.
These additional queries allow the user to ask additional boolean queries of the
system, but a value of \true does not indicate an authorization was granted.
\begin{definition} \label{def:system}
Given access control model \(\model = \tup{\states, \requests}\), an
\emph{access control system} within \model is a state transition system,
\(\system{S} = \tup{\states, \commands, \queries}\), where \(\commands\) is the set of commands, where each command \(\command \in \commands\) is a function \(\states \to \states\), and \(\queries \supseteq \requests\) is
the set of queries, where each query \(\query \in \queries\) is a function
\(\states \to \set{\true, \false}\).
\end{definition}

We use the notation \(\trans\prm{\state, \command}\) to denote the state
resulting from executing \command in \state (that is, \(\command\prm{\state}\)),
and \(terminal\prm{\state, \command_1 \circ \command_n}\) to denote the final
state produced by repeatedly applying \(\trans\) to the commands \(\command_1,
\ldots, \command_n\) starting from state \state:
\(\trans\prm{\ldots\trans\prm{\state, \command_1}, \ldots, \command_n}\).

A system based on the example role-based model must define commands to transform
the state: e.g., to assign roles to users, and assign permissions to roles.
Additional queries beyond the model's requests may include those of the form
``Is user \(u\) a member of role \(r\)?''

Next, we define an access control mapping, which maps one system to another but
does not enforce any simulation properties. We define a mapping as motivated in
\cref{sec:implementable:requirements} so that it can represent any implementable
expressiveness simulation.
\begin{definition} \label{def:mapping}
Given two access control systems, \(\system{S} = \tup{\states^\system{S},
\commands^\system{S}, \queries^\system{S}}\) and \(\system{T} =
\tup{\states^\system{T}, \commands^\system{T}, \queries^\system{T}}\), a
\emph{mapping} from \system{S} to \system{T} is a triple of functions \(\mapping
= \tup{\mapping_\states, \mapping_\commands, \mapping_\queries}\), where:
\begin{itemize}

\item \(\mapping_\states : \states^\system{S} \to \states^\system{T}\) is the
state mapping

\item \(\mapping_\commands : \commands^\system{S} \times \states^\system{T} \to
\prm{\commands^\system{T}}^*\) is the command mapping

\item \(\mapping_\queries = \queries^\system{S} \times \states^\system{T} \to
\set{\true, \false}\) is the query decider

\end{itemize}
\end{definition}

\begin{figure}
\centering
\includegraphics[width=0.4\columnwidth]{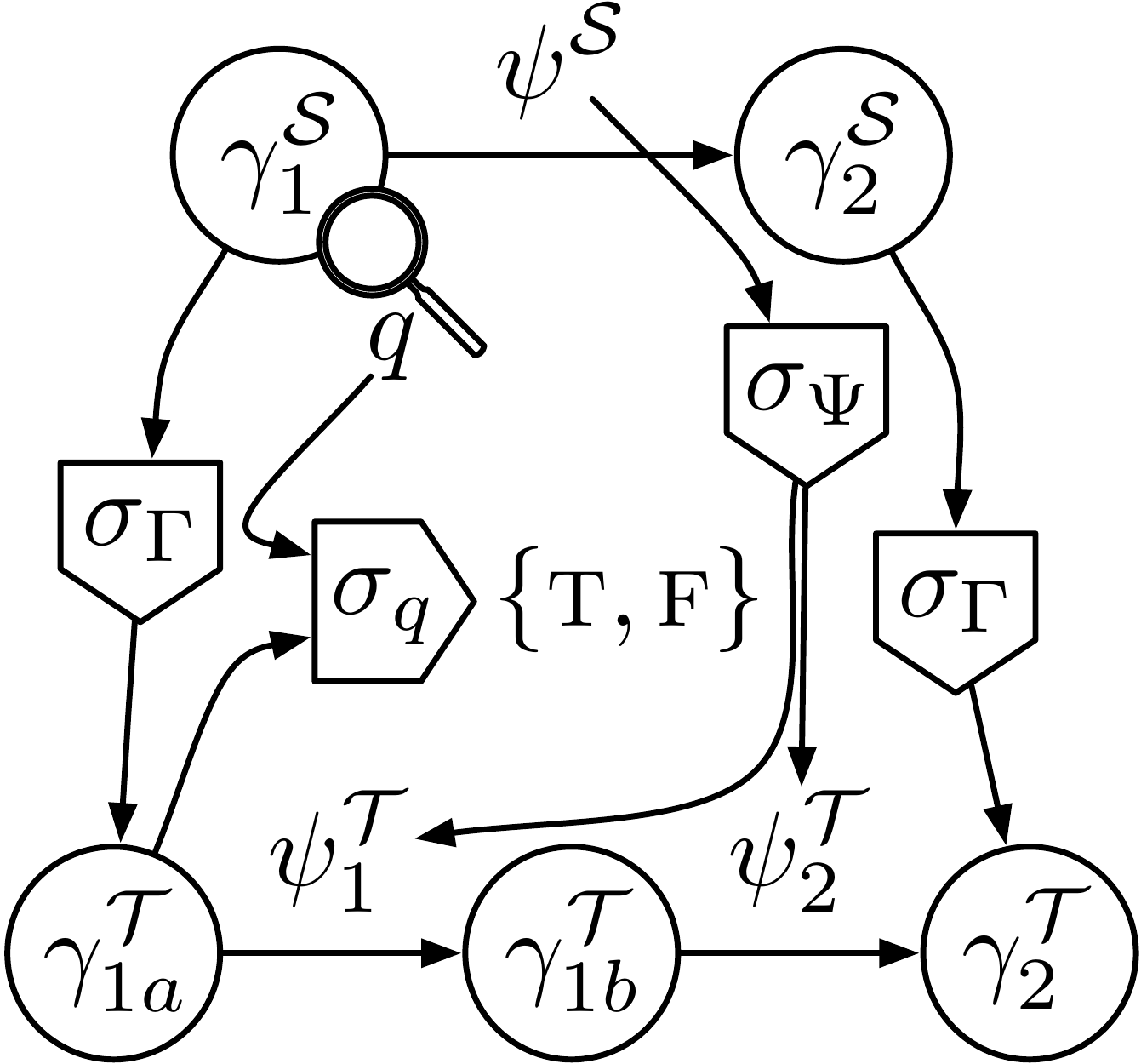}
\caption{The general form of an implementable expressiveness mapping.}
\label{fig:implementable}
\end{figure}

This \lcnamecref{def:mapping} is demonstrated in \cref{fig:implementable}. Each
function takes its most general form that satisfies the requirements from
\cref{sec:implementable:requirements}. Thus, the \lcnamecref{def:mapping}
remains general (it does not enforce any specific security requirements yet),
while ensuring that any such mappings can generate implementable procedures for
using the simulating system in place of the simulated system.

To demonstrate \cref{def:mapping}, consider mapping a simple access control list
system to the role-based system described throughout this
\lcnamecref{sec:implementable}. The state mapping can map each ACL state to a
role-based state in which each user \(u\) has a unique role \(r_u\), and each
user's role is assigned the permissions from the ACL state. The command mapping
can map, e.g., ``grant \(u\) access to \(o\)'' to ``assign \(o\) to role
\(r_u\).'' The query mapping would then map ``Can \(u\) access \(o\)?'' to ``Is
\(u\) authorized for \(o\)?''

%% file: properties.tex
\section{Expressiveness Simulation Properties} \label{sec:properties}

In this \lcnamecref{sec:properties}, we describe the lattice of properties that
we use to taxonomize access control expressiveness simulations.

\subsection{Overview of dimensions of properties}
\label{sec:properties:overview}

\begin{figure}
\null
\hfill
\begin{subfigure}{0.36\columnwidth}
\includegraphics[width=\columnwidth]{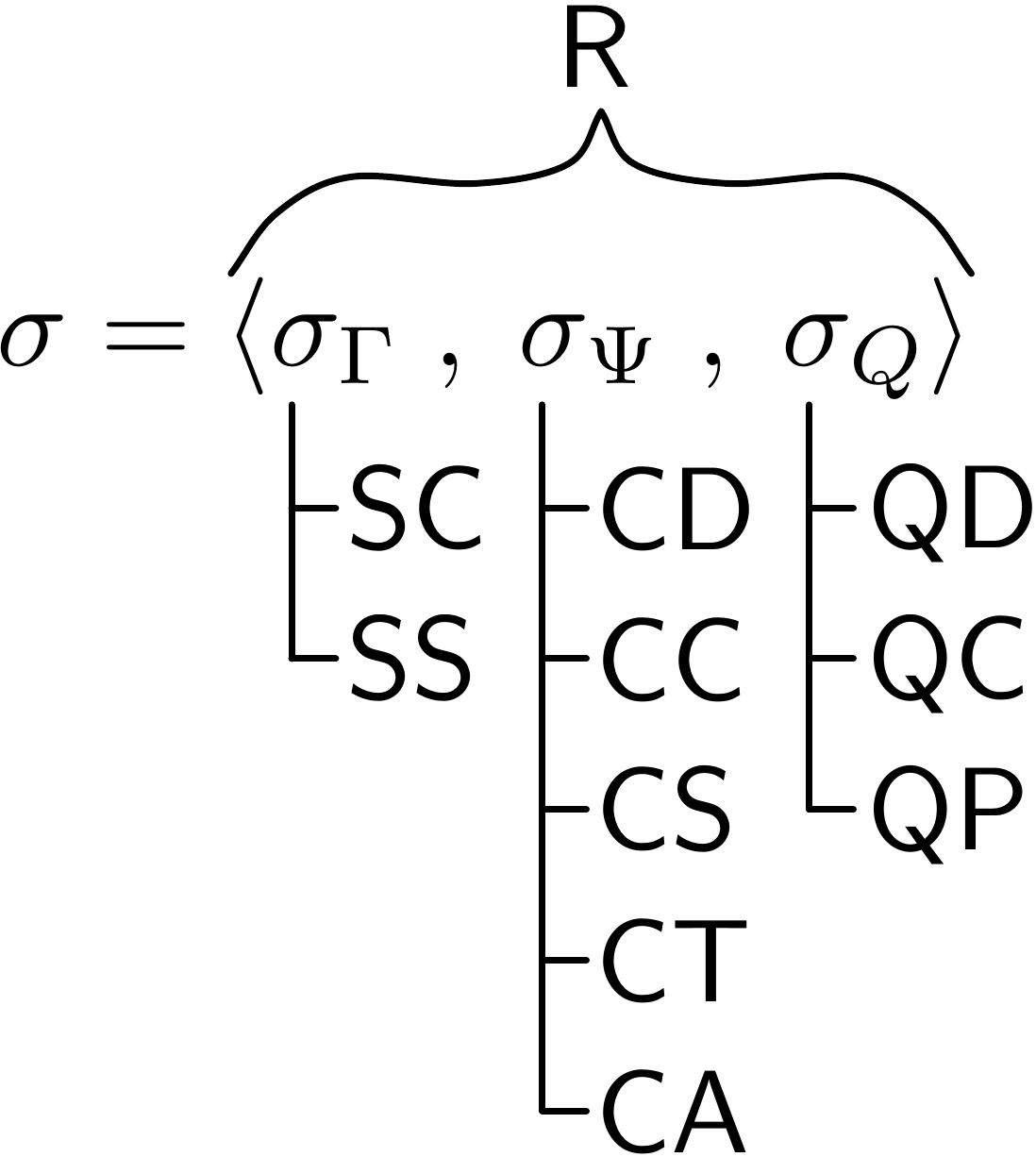}
\end{subfigure}
\hfill
\begin{subfigure}{0.62\columnwidth}
\smaller[1]
\begin{tabular}{c l}
\textbf{Symbol} & \textbf{Description} \\
\propsym{SC} & State correspondence \\
\propsym{SS} & State storage \\
\\
\propsym{CD} & Command mapping dependence \\
\propsym{CC} & Command mapping complexity \\
\propsym{CS} & Command mapping stuttering \\
\propsym{CT} & Trace structure \\
\propsym{CA} & Actor preservation \\
\\
\propsym{QD} & Query decider dependence \\
\propsym{QC} & Query decider complexity \\
\propsym{QP} & Query preservation \\
\\
\propsym{R}  & Reachability
\end{tabular}
\end{subfigure}
\hfill
\null
\caption{An overview of the dimensions of expressiveness simulation properties}
\label{fig:dimensions}
\end{figure}

In order for a mapping to be considered a simulation, it must enforce additional
properties over \cref{def:mapping}. We restrict this definition in a variety of
ways. Although no set of restrictions can be shown to be \emph{the} full,
correct set for all conceivable simulations, there are naturally three
categories of restrictions to consider for simulations, given their structure (a
set of three functions): i.e., refinements to each of the state correspondence,
command mapping, and query decider. We also consider restrictions to the
reachability constraints required (a cross-cutting dimension describing how
these functions must relate to one another). A summary of these dimensions is
depicted in \cref{fig:dimensions}.

Our state correspondence \(\mapping_\states\) can be based on any of a handful
of structural definitions, defined by \propsym{SC} (i.e., what elements do we
inspect to determine whether two states correspond?). Further, \propsym{SS} can
limit the amount of storage the state correspondence uses (e.g., \system{T} must
simulate \system{S} using only a linear amount of additional storage).

The command mapping \(\mapping_\commands\) can be restricted by \propsym{CD} in
what state elements it can use to map commands (e.g., whether it can inspect
arbitrary state elements or only those that are exposed via queries).
\propsym{CC} considers limiting the time-complexity of the command mapping
routine. Since the command mapping returns a \emph{sequence} of commands,
\propsym{CS} can limit the number of commands it can return (e.g., only one, or
constant in the size of the state). We identify \propsym{CT}, a dimension of
concurrency-related trace structure restrictions, as well as \propsym{CA},
requiring the simulation to map \system{S} commands executed by certain types of
users only to \system{T} commands executed by that category of users.

The query decider \(\mapping_\queries\) can also be restricted in a number of
ways. Like the command mapping, we may limit what elements of the state the
decider can inspect when deciding how to answer queries within a specific state
(\propsym{QD}), or the time-complexity of the routine (\propsym{QC}). In some
cases a simulation of \system{S} in \system{T} may be required to map certain
\system{S} queries to specific related queries in \system{T}, most notably
authorization requests (e.g., to answer whether user \(u\) should have
permission \(p\) in \system{S}, \system{T} should simply check whether user
\(u\) has permission \(p\) in its current \system{T} state); this type of
restriction is handled in \propsym{QP}.

Finally, our reachability restrictions \propsym{R} define how these three
functions relate, by allowing us to parameterize whether we require one-way
reachability (\system{T} must be able to transition to states corresponding to
all reachable \system{S} states) or bidirectional reachability (\system{T} also
cannot transition to states that do not correspond to reachable \system{S}
states).

The bare minimum set of these simulation properties that must be enforced for a
mapping to be considered a simulation is a notion of state correspondence and a
reachability relation. We present the definition of implementable expressiveness
simulation, which refines the mapping by enforcing these properties.
\begin{definition} \label{def:implementable}
Given two access control systems, \(\system{S} = \tup{\states^\system{S},
\commands^\system{S}, \queries^\system{S}}\) and \(\system{T} =
\tup{\states^\system{T}, \commands^\system{T}, \queries^\system{T}}\) and a
mapping \(\mapping = \tup{\mapping_\states, \mapping_\commands, \mapping_\queries}\) from
\system{S} to \system{T}, an \emph{implementable expressiveness simulation} of
\system{S} in \system{T} based on \mapping is defined as \(\mapping^\prime = \tup{\mapping_\states,
\mapping_\commands, \mapping_\queries, \thecorrespond, R}\), where:
\begin{itemize}

\item \(\thecorrespond \subseteq \states^\system{S} \times \states^\system{T}\)
is the state correspondence, and \(\forall \state \in \states^\system{S}, \state
\correspond \mapping_\states\prm{\state}\)

\item \(R\) is a reachability restriction

\end{itemize}
\end{definition}

We define all properties over the expressiveness simulation \(\mapping =
\tup{\mapping_\states, \mapping_\commands, \mapping_\queries, \thecorrespond,
R}\). Unless otherwise noted, properties within a dimension are totally ordered
from most to least strict.

\subsection{State correspondence properties} \label{sec:properties:state}

As discussed in \cref{sec:implementable:definition}, the state correspondence of
an implementable simulation of \system{S} in \system{T} is a function,
\(\mapping_\states : \states^\system{S} \to \states^\system{T}\) mapping each
state in \system{S} to a state in \system{T}. There are several ways in which we
can restrict this mapping.

\dimension{SC}{State correspondence structure}

This dimension of properties restricts the way in which corresponding states are
structurally similar. All properties within this dimension were inspired by
state correspondence relations from prior expressiveness simulations; other
application-specific state correspondence relations are conceivable.

\begin{sproperty}[\prop{SCs}{Structure-correspondent}]
\(\state^\system{S} \correspond[\propsym{s}] \state^\system{T} \defas \forall
S_i \in \statedecompose{\state^\system{S}}.(S_i \in
\statedecompose{\state^\system{T}})\)
\end{sproperty}

\begin{sproperty}[\prop{SCq}{Query-correspondent}]
\(\state^\system{S} \correspond[\propsym{q}] \state^\system{T} \defas \forall
\query \in \queries^\system{S}.( \state^\system{S} \entails \query \iff
\mapping_\queries\prm{\query, \state^\system{T}} = \true)\)
\end{sproperty}

\begin{sproperty}[\prop{SCa}{Authorization-correspondent}]
\(\state^\system{S} \correspond[\propsym{a}] \state^\system{T} \defas \forall
\request \in \requests^\system{S}.(\state^\system{S} \entails \request \iff
\mapping_\queries\prm{\request, \state^\system{T}} = \true)\)
\end{sproperty}

\emph{Authorization-correspondent} simulations enforce that every
\(\state^\system{S}\) maps to a \(\state^\system{T}\) that agrees on all
authorization requests: any permission granted/denied in \(\state^\system{S}\)
must also be granted/denied in \(\state^\system{T}\). Requests that exist in
\system{T} but not in \system{S} are not restricted. This type of correspondence
is used in \cite{Sandhu92,SG93,SM98,MS99,OSM00,CDM01}.
\emph{Query-correspondence} requires that \(\state^\system{S}\) and
\(\state^\system{T}\) agree on \emph{all} queries, not just authorization
requests. This type of correspondence is used in the expressiveness simulations
of \cite{HMG13,TL07}.

Finally, \emph{structure-correspondent} simulations require all corresponding
state elements to be identical. If \(\state^\system{S}\) structure-corresponds
to \(\state^\system{T}\), then every set in \(\state^\system{S}\) exists in
\(\state^\system{T}\), and contains all the same elements (\(\state^\system{T}\)
may contain additional sets or relations). Thus, if \(\state^\system{S}\)
contains sets of users and permissions, and a relation between them (a subset of
\(\text{users} \times \text{permissions}\)) specifying accesses,
\(\state^\system{T}\) must contain identical sets of users and permissions, and
an identical set of \(\tup{\text{user}, \text{permission}}\) pairs. This notion
of state correspondence is used in \cite{ALS96}.

The type of state correspondence used is a central characteristic of a type of
simulation. Enforcing a state correspondence that is too weak can allow the
simulating system to diverge from the simulated system in unexpected ways, while
a state correspondence that is too strong will cause the simulating system to
track the simulated system more closely than necessary (e.g., by constraining
the values of queries that the deployment never needs to ask). Thus, choosing a
particular state correspondence is choosing how closely the simulating system
must stay to the simulated system.

\dimension{SS}{State storage}

An orthogonal class of restrictions that can be placed on the state
correspondence relation involve its allowed storage. Here, we restrict the size
of \(\state^\system{T} = \mapping_\states\prm{\state^\system{S}}\) with respect
to \(\state^\system{S}\).

\begin{sproperty}[\prop{SSl}{Linear storage}]
\(\exists c \in \mathbb{R}^+, s \in \mathbb{Z}^+ :
	\forall \state \in \states^\system{S} :
		\card{\state} \geq s \implies
		\card{\mapping_\states\prm{\state}} \leq c \card{\state}
\)
\end{sproperty}

\begin{sproperty}[\prop{SSp}{Polynomial storage}]
\(\exists k \in \mathbb{R}^{+}, s \in \mathbb{Z}^+ :
	\forall \state \in \states^\system{S} :
		\card{\state} \geq s \implies
		\card{\mapping_\states\prm{\state}} \leq \card{\state}^k
\)
\end{sproperty}

\begin{property}[\prop{SS\infty}{Unbounded storage}]
\nullproptext
\end{property}

A \emph{linear storage} simulation says that \(\state^\system{T}\) can grow at
most linearly with \(\state^\system{S}\), while in a \emph{polynomial storage}
simulation, the size of \(\state^\system{T}\) is bounded by a polynomial in the
size of \(\state^\system{S}\). The most obvious result of enforcing properties
within \propsym{SS} is limited trusted storage, but it can also limit iteration
over the resulting state (e.g., if an action must be taken for each document in
the simulating system, \propsym{SSl} ensures that this sequence of actions is
linear in the size of the simulated state).


\subsection{Command mapping properties} \label{sec:properties:command}

Recall that the command mapping for an implementable simulation
(\cref{def:implementable}) is a function \(\mapping_\commands :
\commands^\system{S} \times \states^\system{T} \to
\prm{\commands^\system{T}}^*\) that returns the sequence of \system{T} commands
needed to simulate an \system{S} command starting from a particular \system{T}
state. Thus, it allows us to simulate \system{S} commands in an active
simulation using \system{T}. We now discuss the ways in which we can restrict
this mapping.

\dimension{CD}{Command mapping dependence}

While \cref{def:implementable} maps each \system{S} command and \system{T} state
to a sequence of \system{T} commands, some previous works use more strict
command mappings, mapping each \system{S} command to a sequence of \system{T}
commands without considering the state~\cite{CDM01}. In between these options,
we may map each \system{S} command and \system{T} \emph{theory}, calculating the
sequence of \system{T} commands by observing only the queriable portions of the
\system{T} state. \emph{Command mapping dependence} thus restricts the
information that the command mapping can consider about a \system{T} state when
calculating the trace of \system{T} commands to execute.

\begin{sproperty}[\prop{CDi}{Independent command mapping}]
\(\exists \mapping^\prime: \commands^\system{S} \to
\prm{\commands^\system{T}}^*.(\mapping_\commands\prm{\command, \state} \equiv
\mapping^\prime\prm{\command})\)
\end{sproperty}

\begin{sproperty}[\prop{CDt}{Theory-dependent command mapping}]
\(\exists \mapping^\prime: \commands^\system{S} \times \theory{\system{T}} \to
\prm{\commands^\system{T}}^*.(\mapping_\commands\prm{\command, \state} \equiv
\mapping^\prime\prm{\command, \theory{\state}})\)
\end{sproperty}

\begin{property}[\prop{CDs}{State-dependent command mapping}]
\nullproptext
\end{property}

With \emph{independent command} simulations, \system{S} commands must be
precompiled to \system{T} commands which will work in any reachable \system{T}
state. This is a restriction placed by \cite{CDM01}. \emph{Theory dependent}
command mappings allow limited inspection of the \system{T} state; this
restriction allows the sequence of \system{T} commands to be determined based
only on the \emph{theory} of the \system{T} state: the values of all \system{T}
queries in the state. If two \system{T} states answer all queries the same way,
the same \system{T} commands would be used in both to simulate an \system{S}
command. With this restriction, the monitor that transforms \system{S} inputs
into \system{T} procedures need not be more privileged than users of the access
control system, since queries are the user's only API to observe the state.

Finally, \emph{state-dependent} command mappings can arbitrarily observe the
state. This requires a monitor that is privileged enough to observe elements of
the state that are not queriable, and two states that answer all queries
identically may simulate commands differently depending on unobservable state.

\dimension{CC}{Command mapping complexity}

Having considered the inputs available to the command mapping, we now consider
the time complexity of this mapping. Note that this is measured as the increase
in time as the \emph{state} grows and thus is meaningless for independent
command.

\begin{property}[\prop{CCc}{Constant command mapping}]
\(\forall \command \in \commands^\system{S}\), the algorithm for
\(\mapping_\command\prm{\state} = \mapping_\commands\prm{\command, \state}\) has
time complexity \(T\prm{n} \in \bigO{1}\)
\end{property}

\begin{property}[\prop{CCl}{Linear command mapping}]
\(\forall \command \in \commands^\system{S}\), the algorithm for
\(\mapping_\command\prm{\state} = \mapping_\commands\prm{\command, \state}\) has
time complexity \(T\prm{n} \in \bigO{n}\)
\end{property}

\begin{property}[\prop{CC\infty}{Unbounded command mapping}]
\nullproptext
\end{property}

\emph{Constant command} simulations do not allow more processing time for bigger
states. Thus, the command mapping cannot loop over sets within the state. With
\emph{linear command}, the command mapping can take time linear in the size of
the state, e.g., looping over sets in the state, but cannot contain double loops
over sets, sort sets, etc. Finally, \emph{unbounded command} simulations put no
limit on the complexity of the command mapping (though we may expect it to have
to be tractable, e.g., poly-time).

\dimension{CS}{Command mapping stuttering}

Since the command mapping maps an \system{S} state to a sequence of \system{T}
states, we may restrict the number of commands that can be used to simulate a
single \system{S} command.

\begin{property}[\prop{CS1}{Lock-step}]
\(\forall \command \in \commands^\system{S}, \state \in \states^\system{T} :
\card{\mapping_\commands\prm{\command, \state}} \leq 1\)
\end{property}

\begin{property}[\prop{CSc}{Constant step}]
\(\exists c: \forall \command \in \commands^\system{S}, \state
\in \states^\system{T} : \card{\mapping_\commands\prm{\command, \state}} \leq
c\)
\end{property}

\begin{property}[\prop{CS\infty}{Unbounded step}]
\nullproptext
\end{property}

A \emph{lock-step} simulation allows at most one \system{T} command for each
simulated \system{S} command. This mitigates concurrency issues for multiuser
systems, since the system does not pass through potentially inconsistent states
between command executions. \emph{Constant step} simulations allow multiple
commands to be used, but only a number constant in the size of the state. Thus,
multiple actions can be taken, but not, e.g., a command for each user in the
system. Finally, \emph{unbounded step} does not restrict how many \system{T}
commands can be executed per \system{S} command.

\dimension{CT}{Trace structure}

This class of properties enforces structural constraints on the traces of
commands returned by the command mapping. This can address the potentially
inconsistent states between start and end states in traces generated by the
command mapping. Here, we present several examples of trace restrictions, using
the notation \(terminal\prm{\state, \command_1, \cdots, \command_j}\) to denote
the end state resulting from executing the sequence of commands \(\command_1,
\cdots, \command_j\), starting from the state \state. Note that this dimension
of properties is not totally ordered.

{ 

\newcommand{\shim}{\hskip1.5em\relax}

\begin{property}[\prop{CT1}{Semantic lock-step}]
\begin{smallign*}
& \forall \command \in \commands^\system{S}, \state^\system{S} \in \states^\system{S}, \state^\system{T} \in \states^\system{T}.(\\
&\shim \exists \overbar{\command} = \tup{\command_1, \command_2, \ldots, \command_m} \in \prm{\commands^\system{T}}^*, i \in \eirange{1, m}.(\\
&\shim\shim \mapping_\commands\prm{\command, \state^\system{T}} = \overbar{\command} \land {}\\
&\shim\shim \forall j \in \ierange{1, i}.(\state^\system{S} \correspond \state^\system{T} \implies {}\\
&\shim\shim\shim \state^\system{S} \correspond terminal\prm{\state^\system{T}, \command_1 \cdots \command_j}) \land {}\\
&\shim\shim \forall j \in \iirange{i, m}.(\state^\system{S} \correspond \state^\system{T} \implies {}\\
&\shim\shim\shim \trans\prm{\state^\system{S}, \command} \correspond terminal\prm{\state^\system{T}, \command_1 \cdots \command_j}) ) )
\end{smallign*}
\end{property}

First, a \emph{semantic lock-step} simulation can appear to be lock-step (i.e.,
it does not enter any inconsistent states), because even though it is allowed to
execute multiple \system{T} commands to simulate a single \system{S} command,
only one of those commands is allowed to make correspondence-related changes.
That is, consider the sequence of \system{T} states constructed by executing the
sequence of commands \(\mapping_\commands\prm{\command^\system{S},
\state^\system{T}}\). In semantic lock-step, all of these states must correspond
to the either the start state in \system{S} or the end state in \system{S}, and
once the transition from start state to end state is made, the remaining states
must all be equivalent to the end state. Thus, from the point of view of a user
who can ask any combination of queries, the simulation appears to be lock-step.
This restriction is depicted in \cref{fig:semanticlock}.

\begin{figure}
\centering
\includegraphics[width=0.65\columnwidth]{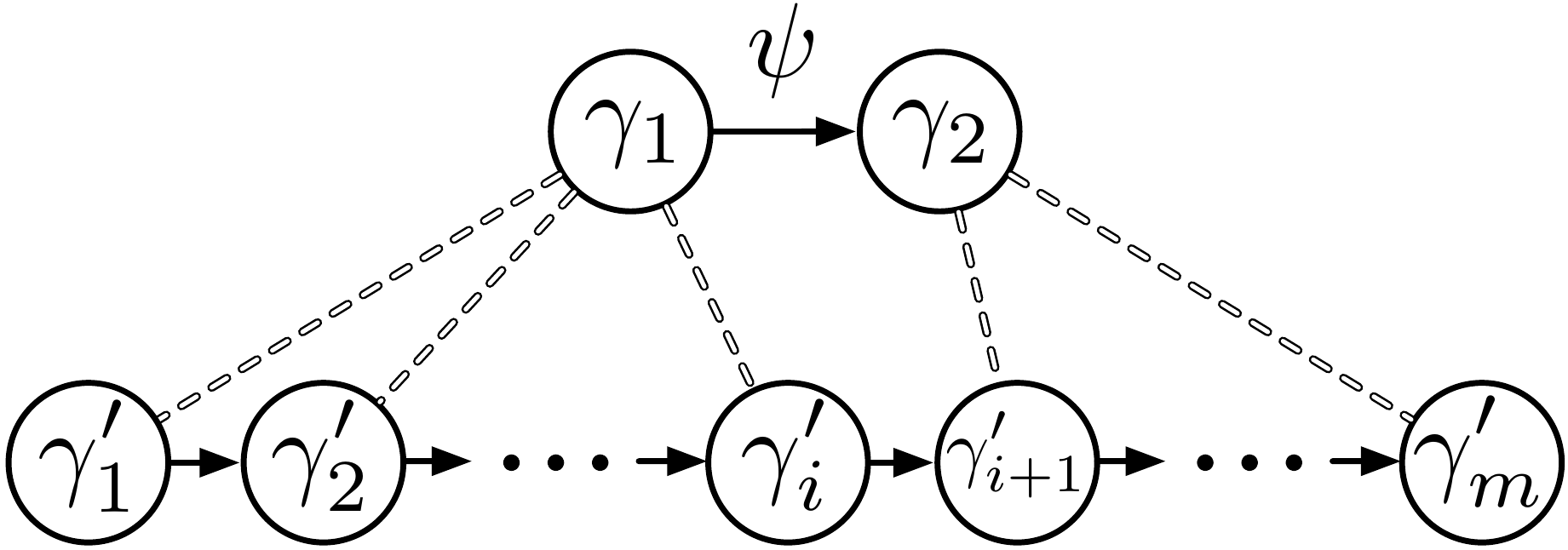}
\caption{A graphical representation of semantic lock-step}
\label{fig:semanticlock}
\end{figure}

\begin{sproperty}[\prop{CTq}{Query monotonic}]
{\smaller[2]
\(\forall \command \in \commands^\system{S}, \state \in \states^\system{T}, \query \in \queries^\system{T}.monotonic\prm{\command, \state, \query}\), where:
}
\begin{smallign*}
&monotonic\prm{\command, \state, \query} \triangleq \exists \overbar{\command} = \tup{\command_1, \command_2, \ldots, \command_m} \in \prm{\commands^\system{T}}^*.(\\
&\shim \mapping_\commands\prm{\command, \state} = \overbar{\command} \land {}\\
&\shim \forall i \in \eerange{1,m}.(\\
&\shim\shim terminal\prm{\state, \command_1 \cdots \command_i} \entails \query \implies {}\\
&\shim\shim\shim (terminal\prm{\state, \command_1 \cdots \command_{i-1}} \entails \query \lor terminal\prm{\state, \command_1 \cdots \command_m} \entails \query) \land {}\\
&\shim\shim terminal\prm{\state, \command_1 \cdots \command_i} \nentails \query \implies {}\\
&\shim\shim\shim (terminal\prm{\state, \command_1 \cdots \command_{i-1}} \nentails \query \lor terminal\prm{\state, \command_1 \cdots \command_m} \nentails \query) ) )
\end{smallign*}
\end{sproperty}

Consider the start and end states of a trace in \system{T}, \state and
\(\state^\prime\), respectively. Let \(\queries^{+}\) be the set of queries that
become true in \(\state^\prime\) that were false in \state, and \(\queries^{-}\)
be the set of queries that become false in \(\state^\prime\) that were true in
\state. During the trace from \state to \(\state^\prime\), \emph{query
monotonicity} enforces that no queries are made true except \(\queries^{+}\),
and no queries are made false except \(\queries^{-}\). Thus, from the point of
view of a user who can ask only single queries, the simulation appears to be
lock-step.

\begin{sproperty}[\prop{CTa}{Access monotonic}]
\(\forall \command \in \commands^\system{S}, \state \in \states^\system{T}, \request \in \requests^\system{T}.monotonic\prm{\command, \state, \request}\)
%
%
\end{sproperty}

\emph{Access monotonicity} is similar to query monotonicity but considering only
authorization requests. Let \(\requests^{+}\) be the set of requests that become
allowed in \(\state^\prime\) that were denied in \state, and \(\requests^{-}\)
be the set of requests that become denied in \(\state^\prime\) that were allowed
in \state. During the trace from \state to \(\state^\prime\), access
monotonicity enforces that no requests are granted except \(\requests^{+}\), and
no requests are revoked except \(\requests^{-}\).

\begin{property}[\prop{CTs}{Non-contaminating}]
\begin{smallign*}
& \forall \command \in \commands^\system{S}, \state^\system{T} \in \states^\system{T}.(\\
&\shim \exists \overbar{\command} = \tup{\command_1, \command_2, \ldots, \command_m} \in \prm{\commands^\system{T}}^*.(\\
&\shim\shim \mapping_\commands\prm{\command, \state^\system{T}} = \overbar{\command} \land {}\\
&\shim\shim \forall \state^\system{T}_i \in \set{\state^\system{T}_i \mid \exists \command_i \in \overbar{\command} : \state^\system{T}_i = terminal\prm{\state^\system{T}, \command_1 \cdots \command_i}}.(\\
&\shim\shim\shim Allowed\prm{\state^\system{T}_i} \subseteq Allowed\prm{\state^\system{T}} \lor {}\\
&\shim\shim\shim Allowed\prm{\state^\system{T}_i} \subseteq Allowed\prm{terminal\prm{\state^\system{T}, \overbar{\command}}} ) ) )
\end{smallign*}
\end{property}

The \emph{non-contaminating} trace property ensures that no two accesses are
allowed in the same state that are not both allowed in either the start or end
state. This prevents, e.g., an intermediate state where a file can be accessed
by two users simultaneously when simulating a command intended to \emph{switch}
which user can access the file. This definition uses the \(Allowed\prm{\state}\)
notation, indicating the set of all permissions \(p\) allowed in state \state
(i.e., such that \(\state \entails p\)).

\dimension{CA}{Actor preservation}

Actor preservation properties restrict which users can be invoked in \system{T}
to handle \system{S} commands. Here, we assume that \(\alpha\prm{\command}\)
denotes the actor executing the command \command. Note that this requires system
support (e.g., the executing actor being an implicit argument passed to a
command) in order for a simulation to be executable.

\begin{property}[\prop{CA\top}{Self-execution}]
\(\forall \command^\system{S} \in \commands^\system{S}, \state \in
\states^\system{T}, \forall \command^\system{T} \in
\mapping_\commands\prm{\command^\system{S}, \state},
\alpha\prm{\command^\system{S}} = \alpha\prm{\command^\system{T}}\)
\end{property}

\begin{property}[\prop{CAa}{Administration-preservation}]
Let \(A\) be the administrative subset of executing entities in the system.
\(\forall \command^\system{S} \in \commands^\system{S}, \state \in
\states^\system{T}, \forall \command^\system{T} \in
\mapping_\commands\prm{\command^\system{S}, \state},
\alpha\prm{\command^\system{T}} \in A \implies \alpha\prm{\command^\system{S}}
\in A\)
\end{property}

\emph{Self-execution} says that any command in \system{S} executed by any user
\(u\) must be mapped to a sequence of commands in \system{T}, all of which are
executed by \(u\).
\emph{Administration-preservation} prevents the invocation of
administrators in \system{T} where they were not needed in \system{S}. In an
administration-preserving simulation, any command in \system{S} executed by a
non-administrative user is mapped to a sequence of commands in \system{T}, none
of which is executed by an administrator. Other forms of actor preservation, as
well as defining the set of administrators, are application-specific.

\subsection{Query decider properties} \label{sec:properties:query}

We defer the bulk of the technical discussion of the query decider restrictions
to \inlongshort{\cref{sec:queryproperties}}{the technical report accompanying
this paper~\cite{csf2015extended}}, as they are largely similar to the command
mapping restrictions. Query decider dependence (\propsym{QD}), like command
mapping dependence (\propsym{CD}), restricts the information that the query
decider can consider about a \system{T} state when deciding the truth value of
an \system{S} query in that state. Query decider complexity (\propsym{QC})
restricts the runtime of the routine.

Query preservation (\propsym{QP}) indicates which queries need to stay the same
as they are mapped from system \system{S} to system \system{T}. A particular
application may require any given set of queries to be preserved; the most
common property in this dimension is \emph{authorization preservation}, which
enforces that the query decider maps each \system{S} request to the value of the
identical request in the \system{T} state. This can be seen as ensuring that
\system{T} is using its model ``as intended'' (i.e., forcing it to answer
simulated requests as it would its own native requests).

\subsection{Reachability} \label{sec:properties:reachability}

\dimension{R}{Reachability}

The last dimension of properties we consider ties the mappings together to
ensure the simulation is indeed what one could consider a simulation in the
classic sense. A state correspondence, query decider, and command mapping do not
automatically define a simulation without reachability constraints. Here, we
define forward and bidirectional reachability, two variants of this type of
constraint (note that these properties are presented in \emph{increasing}
strictness since the latter builds upon the former).

\begin{property}[\prop{R\reach}{Forward reachability}]
\begin{smallign*}
& \forall \state^\system{S}_0, \state^\system{S}_1 \in \states^\system{S}, \state^\system{T}_0 \in \states^\system{T}.( \\
&\shim \state^\system{S}_0 \correspond \state^\system{T}_0 \land \state^\system{S}_0 \mapsto \state^\system{S}_1 \implies \exists \state^\system{T}_1 \in \states^\system{T}.( \\
&\shim\shim \state^\system{T}_0 \Mapsto \state^\system{T}_1 \land \state^\system{S}_1 \correspond \state^\system{T}_1 ) )
\end{smallign*}
\end{property}

\begin{property}[\prop{R\bireach}{Bidirectional reachability}] Forward
reachability, and:
\begin{smallign*}
& \forall \state^\system{S}_0 \in \states^\system{S}, \state^\system{T}_0, \state^\system{T}_1 \in \states^\system{T}.( \\
&\shim \state^\system{S}_0 \correspond \state^\system{T}_0 \land \state^\system{T}_0 \mapsto \state^\system{T}_1 \implies \exists \state^\system{S}_1 \in \states^\system{S}.( \\
&\shim\shim \state^\system{S}_0 \Mapsto \state^\system{S}_1 \land \state^\system{S}_1 \correspond \state^\system{T}_1 ) )
\end{smallign*}
\end{property}

In \emph{forward reachability}, any transition made in \system{S} must be
possible in \system{T}. If \(\state^\system{S}_0\) corresponds to
\(\state^\system{T}_0\), and \(\state^\system{S}_1\) can be reached from
\(\state^\system{S}_0\) via the commands of \system{S}, then
\(\state^\system{S}_1\) must correspond to a state \(\state^\system{T}_1\) in
\system{T} that is reachable from \(\state^\system{T}_0\). The notion of state
correspondence is determined by the property chosen in dimension \propsym{SC}.

\emph{Bidirectional reachability} (or bi-reachability), also requries that
\system{T} cannot enter a state that does not correspond to a reachable state in
\system{S}. If \(\state^\system{S}_0\) corresponds to \(\state^\system{T}_0\),
and \(\state^\system{T}_1\) is reachable from \(\state^\system{T}_0\) by
executing a command, then there must exist an \system{S} state
\(\state^\system{S}_1\) that corresponds to \(\state^\system{T}_1\) and that is
reachable from \(\state^\system{S}_0\) by executing one or more commands. This
process may make use of multiple steps, since the procedure for finding the
corresponding \system{S} states does not need to be constructed, these states
must simply exist. The operational advantage of enforcing \propsym{R\bireach} is
that, even if the simulating system's native operations are exposed to users,
the system can never enter a state that does not have an equivalent in the
simulated system.

} 

%% file: positioning.tex
\section{Positioning Existing Simulations}
\label{sec:positioning}

As mentioned in \cref{sec:properties:overview}, no set of properties can be
proven to describe all conceivable simulations. In this
\lcnamecref{sec:positioning}, we support the set of properties defined in this
work by showing that it can \emph{precisely} describe the wide range of existing
expressiveness simulations.

\subsection{Expressiveness using Simulation Properties}

We will now draw the formal distinction between a simulation and expressiveness.
Here, we use \simulates[\simulation]{\system{T}}{\system{S}} to denote,
``\system{T} can admit a simulation of type \simulation of \system{S},'' and
\reduction[\simulation]{\system{S}}{\system{T}} to denote, ``\system{T} is at
least as expressive as \system{S} with respect to simulations of type
\simulation.''

While previous work considers the expressiveness result to be equivalent to a
simulation (i.e., \(\simulates[\simulation]{\system{T}}{\system{S}} \equiv
\reduction[\simulation]{\system{S}}{\system{T}}\)), expressiveness in a
practical sense is subject to a subtle distinction. Since we mean for
expressiveness to be \emph{implementable} (i.e., if \system{T} is as expressive
as \system{S}, then \system{T} can be used in place of \system{S}),
expressiveness within the domain of simulation properties should mean the
following: if \system{T} is as expressive as \system{S}, then \system{T} can
simulate any system that \system{S} can simulate. Thus, we define
\emph{expressiveness} in the context of a set of simulation properties.
\begin{definition}[Expressiveness] \label{def:expressiveness}
Given access control systems \system{S} and \system{T} and a set of simulation
properties \properties, we say that \system{T} is \emph{at least as expressive}
as \system{S} with respect to \properties (denoted
\reduction[\properties]{\system{S}}{\system{T}}) to mean that, for every system
\system{U}, if \system{S} can simulate \system{U} while enforcing \properties,
then \system{T} can simulate \system{U} while enforcing \properties (\(\forall
\system{U}: \simulates[\properties]{\system{S}}{\system{U}} \implies
\simulates[\properties]{\system{T}}{\system{U}}\)).
\end{definition}

We first point out that this definition of expressiveness is strictly more
general than the more traditional (often implied) notion. Since \system{S} can
trivially simulate itself, \reduction[\simulation]{\system{S}}{\system{T}}
implies \simulates[\simulation]{\system{T}}{\system{S}}. The additional
generalization can be viewed from a formal standpoint as dropping the
(incorrect) assumption that all types of simulation are transitive (i.e., that
\simulates{\system{T}}{\system{S}} and \simulates{\system{S}}{\system{U}} imply
\simulates{\system{T}}{\system{U}}). For instance, assume that \system{T} can
simulate \system{S} and \system{S} can simulate \system{U}, each with a
quadratic increase in state storage. While \system{T} may be able to simulate
\system{U}, this simulation may require greater than quadratic storage.

From a more intuitive standpoint, we point out that, except in the case of
custom-built access control solutions, any deployment is a simulation of a
workload (i.e., ideal operation) using an existing system. That is, unless
\system{S} is custom-made to exactly satisfy the desired workload, replacing it
with \system{T} is not a matter of whether \system{T} can simulate \system{S},
but whether \system{T} can admit an equally good simulation of the (perhaps not
formally specified) workload that \system{S} is known to simulate. This concept
is discussed by Kane and Browne~\cite{kane06}, who point out that an access
control implementation is often only an approximation of the desired policy. In
particular, as policy languages get more complex, deployments often make use of
approximations that are easier to analyze and more efficient to enforce than the
overly-expressive policy language.

\subsection{Decomposing Expressiveness Simulations to Properties}
\label{sec:positioning:decomposition}

In order to use the set of expressiveness simulation properties detailed in
\cref{sec:properties} to systematically compare previously proposed notions of
simulation, we present our formal way of stating that a notion of simulation and
a set of simulation properties are equivalent. We call this correspondence
\emph{simulation decomposition}: when a notion of simulation \simulation can be
\emph{decomposed to} a set of simulation properties \properties, then analyses
using \simulation and \properties yield equivalent expressiveness results.
\begin{definition}[Simulation Decomposition] \label{def:decomposition}
Given a notion of access control simulation \simulation and a set of simulation
properties \properties, \simulation can be \emph{decomposed} to \properties
(denoted \decomposes{\simulation}{\properties}) if and only if, for all systems
\system{S} and \system{T}, \(\simulates[\simulation]{\system{T}}{\system{S}}
\iff \reduction[\properties]{\system{S}}{\system{T}}\).
That is, \system{T} admits an \simulation simulation of \system{S} if and only
if \system{T} is at least as expressive as \system{S} with respect to properties
\properties.
\end{definition}

Recall from \cref{def:expressiveness} that
\reduction[\properties]{\system{S}}{\system{T}} says that any system that can be
simulated by \system{S} while preserving properties \properties can can also
simulated by \system{T} while preserving \properties. In light of this, we will
position an existing notion of simulation, \simulation, within the lattice
formed by our simulation properties (i.e., prove
\decomposes{\simulation}{\properties}) by proving the following for the set of
properties \properties:
\begin{enumerate}

\item (\emph{Only-if direction})
\(\simulates[\simulation]{\system{T}}{\system{S}} \land
\simulates[\properties]{\system{S}}{\system{U}} \implies
\simulates[\properties]{\system{T}}{\system{U}}\)

\item (\emph{If direction}) \(\reduction[\properties]{\system{S}}{\system{T}}
\implies \simulates[\simulation]{\system{T}}{\system{S}}\)

\end{enumerate}
We give an example of such a proof in the following
\lcnamecref{sec:positioning:example}.

\subsection{Example Decomposition} \label{sec:positioning:example}

To demonstrate how simulation decomposition proofs are written, we now consider
the Ammann-Lipton-Sandhu simulation~\cite{ALS96}. The ALS simulation considers
access control states as graphs: sets of primitive objects are node types, and
sets of relations are edge types. The set of node types and edge types in the
states of system \system{S} are denoted \(\mathop{NT}\prm{\system{S}}\) and
\(\mathop{ET}\prm{\system{S}}\), respectively. The ALS state correspondence is
then defined as follows (reworded slightly from \cite{ALS96}).
\begin{definition} \label{def:alscorrespond}
A state in system \system{S}, a simulated system, and a state in system
\system{T}, a simulating system, \emph{correspond} iff the graph defining the
state in \system{S} is identical to the subgraph obtained by taking the state in
\system{T} and discarding all nodes (edges) not in
\(\mathop{NT}\prm{\system{S}}\) (\(\mathop{ET}\prm{\system{S}}\)).
\end{definition}

The ALS simulation is defined with respect to this state correspondence.
\begin{definition} \label{def:alssimulation}
Under the definition of correspondence in \cref{def:alscorrespond}, system
\system{T} \emph{simulates} system \system{S} iff the following conditions hold:
\begin{enumerate}

\item If system \system{S} can reach a given state, system \system{T} can reach
a corresponding state.

\item If system \system{T} can reach a given state, system \system{S} can reach
a corresponding state.

\end{enumerate}
\end{definition}

\renewcommand{\propertyset}{\ensuremath{\set{\propsym{SCs}, \propsym{QPa}, \propsym{R\bireach}}}\xspace}

We will now demonstrate the two-step simulation decomposition proof technique
described in \cref{sec:positioning:decomposition} for the ALS simulation.
For the purposes of this proof, let the set of simulation properties
\(\properties = \propertyset\). Recall that \propsym{SCs} is structure state
correspondence, which says that the simulating state must include all of the
unaltered sets from the simulated state; \propsym{QPa} is authorization
preservation, which says that each authorization request must be mapped
identically from simulated to simulating system (and thus the simulating system
must support the same set of requests as the simulated system); and
\propsym{R\bireach} is bireachability, which says that the simulating system can
reach a state which corresponds to each reachable simulated state, and cannot
reach a state which does not correspond to a reachable state in the simulated
system.

\begin{figure*}
\null
\hfill
\begin{subfigure}{0.66\columnwidth}
\centering
{\smaller[1]
\begin{tabular}{r l}
Simulation & Decomposition \\\hline
ALS        & \propsym{SCs\ QPa\ R\bireach} \\
CDMw       & \propsym{SCa\ QPa\ CDi\ R\reach} \\
CDMs       & \propsym{SCa\ QPa\ CDi\ CS1\ R\reach} \\
Ganta      & \propsym{SCa\ QPa\ CTs\ R\bireach} \\
HMG+       & \propsym{SCq\ QDt\ R\reach} \\
HMG+a      & \phantom{\propsym{SCa}} \propsym{QPa} \\
HMG+s      & \phantom{\propsym{SCa}} \propsym{CTa} \\
HMG+p      & \phantom{\propsym{SCa}} \propsym{CAa} \\
SMG        & \propsym{SCa\ R\reach} \\
TL-SMR     & \propsym{SCq\ QD1\ R\bireach}
\end{tabular}}
\caption{Decompositions of surveyed simulations}
\label{fig:positioning}
\end{subfigure}
\hfill
\begin{subfigure}{0.46\columnwidth}
\centering
\includegraphics[width=\columnwidth]{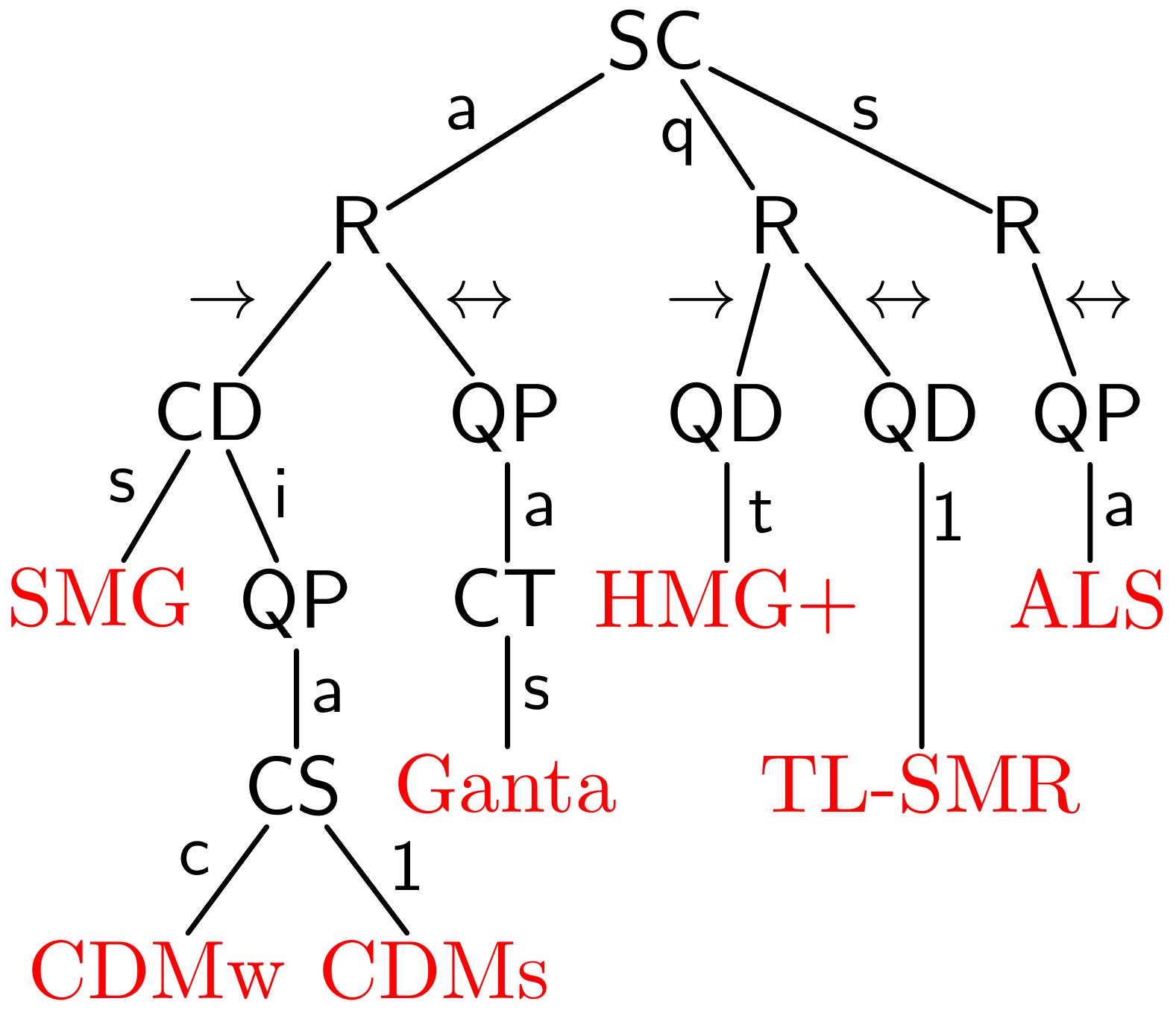}
\caption{Taxonomy of simulations}
\label{fig:taxonomy}
\end{subfigure}
\hfill
\begin{subfigure}{0.5\columnwidth}
\centering
\includegraphics[width=0.66\columnwidth]{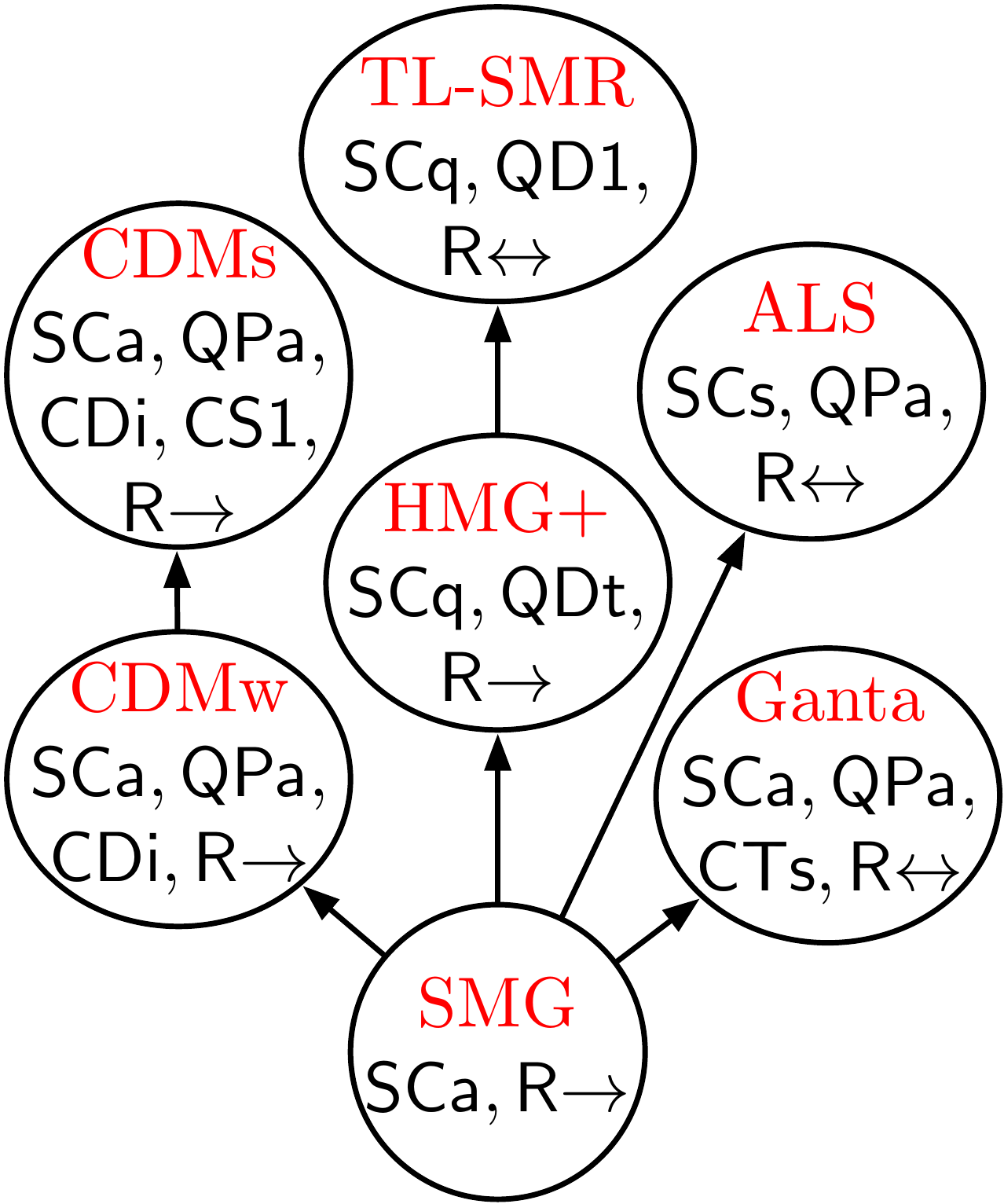}
\caption{Partial lattice of simulations}
\label{fig:partlattice}
\end{subfigure}
\hfill
\null
\caption{}
\end{figure*}

We will demonstrate the two steps of the proof technique by proving two
requesite lemmas. First, step~1 (only-if direction):
\begin{lemma} \label{thm:als:onlyif}
Given access control systems \system{S}, \system{T}, and \system{U},
\[\simulates[ALS]{\system{T}}{\system{S}} \land
\simulates[\properties]{\system{S}}{\system{U}} \implies
\simulates[\properties]{\system{T}}{\system{U}}\]
That is, if \system{T} admits an ALS simulation of \system{S}, and \system{S}
admits a simulation of \system{U} with properties \propertyset, then \system{T}
admits a simulation of \system{U} with properties \propertyset.
\end{lemma}

\begin{proof}
To prove this \lcnamecref{thm:als:onlyif}, we let \system{S}, \system{T}, and
\system{U} be access control systems such that
\simulates[ALS]{\system{T}}{\system{S}} and
\simulates[\properties]{\system{S}}{\system{U}} but are otherwise arbitrary, and
we show that \simulates[\properties]{\system{T}}{\system{U}}.


Choose an arbitrary state \(\state^\system{U}_0 \in \states^\system{U}\) and
command \(\command^\system{U} \in \commands^\system{U}\), and let
\(\trans\prm{\state^\system{U}_0, \command^\system{U}} = \state^\system{U}_1\).
Let \(\state^\system{S}_0 \in \states^\system{S}\) such that
\(\state^\system{U}_0 \correspond[s] \state^\system{S}_0\). Since
\simulates[\properties]{\system{S}}{\system{U}},
\[\exists \state^\system{S}_1 \in
\states^\system{S}.(terminal\prm{\state^\system{S}_0,
\mapping_\commands\prm{\command^\system{U}, \state^\system{S}_0}} =
\state^\system{S}_1 \land \state^\system{U}_1 \correspond[s]
\state^\system{S}_1)\]

Let \(\state^\system{T}_0 \in \states^\system{T}\) such that
\(\state^\system{S}_0 \correspond[s] \state^\system{T}_0\). Since
\simulates[ALS]{\system{T}}{\system{S}},
\[\exists \state^\system{T}_1 \in \states^\system{T}.(\state^\system{T}_0
\Mapsto \state^\system{T}_1 \land \state^\system{S}_1 \correspond[s]
\state^\system{T}_1)\]
Thus, there exists a sequence of \system{T} commands \(\commands^\system{T}_0\)
such that \(terminal\prm{\state^\system{T}_0, \commands^\system{T}_0} =
\state^\system{T}_1\). Define \(\mapping_\commands: \commands^\system{U} \times
\states^\system{T} \to \prm{\commands^\system{T}}^*\) such that it returns
\(\commands^\system{T}_0\) for \(\state^\system{T}_0, \command^\system{U}\).

Then, given \(\state^\system{U}_0, \state^\system{U}_1 \in \states^\system{U},
\state^\system{T}_0 \in \states^\system{T}, \command^\system{U} \in
\commands^\system{U}\) such that \(\trans\prm{\state^\system{U}_0,
\command^\system{U}} = \state^\system{U}_1\), and \(\state^\system{U}_0
\correspond[s] \state^\system{T}_0\),
\[\exists \state^\system{T}_1 \in
\states^\system{T}.(terminal\prm{\state^\system{T}_0,
\mapping_\commands\prm{\command, \state^\system{T}_0}} = \state^\system{T}_1
\land \state^\system{S}_1 \correspond[s] \state^\system{T}_1)\]

Hence, \simulates[\set{\propsym{SCs},
\propsym{R\reach}}]{\system{T}}{\system{U}}. Next, we show \propsym{QPa}.


Choose some arbitrary request \(\request^\system{U}_0 \in \requests^\system{U}\)
and state \(\state^\system{T}_0 \in \states^\system{T}\).

Since \simulates[\properties]{\system{S}}{\system{U}},
\[\forall \request^\system{U} \in \requests^\system{U}, \state^\system{S} \in
\states^\system{S}, \mapping_\queries\prm{\request^\system{U},
\state^\system{S}} = \state^\system{S} \entails \request^\system{U}\]

Thus, we know that \system{S} supports all \system{U} requests, and
corresponding \system{S} and \system{U} states will answer \system{U} requests
identically. Therefore, \(\request^\system{U}_0 \in \requests^\system{S}\).
Since \simulates[ALS]{\system{T}}{\system{S}},
\[\forall \request^\system{S} \in \requests^\system{S}, \state^\system{T} \in
\states^\system{T}, \mapping_\queries\prm{\request^\system{S},
\state^\system{T}} = \state^\system{T} \entails \request^\system{S}\]

Thus, \(\mapping_\queries\prm{\request^\system{U}_0, \state^\system{T}} =
\state^\system{T} \entails \request^\system{U}_0\).


Hence, \simulates[\set{\propsym{SCs}, \propsym{QPa},
\propsym{R\reach}}]{\system{T}}{\system{U}}. Next, we show \propsym{R\bireach}.


Choose some arbitrary states \(\state^\system{T}_0, \state^\system{T}_1 \in
\states^\system{T}\) such that \(\state^\system{T}_0 \mapsto
\state^\system{T}_1\). Let \(\state^\system{S}_0 \in \states^\system{S}\) such
that \(\state^\system{S}_0 \correspond[s] \state^\system{T}_0\). Since
\simulates[ALS]{\system{T}}{\system{S}},
\[\exists \state^\system{S}_1.(\state^\system{S}_0 \Mapsto \state^\system{S}_1
\land \state^\system{S}_1 \correspond[s] \state^\system{T}_1)\]

Let \(\state^\system{U}_0 \in \states^\system{U}\) such that
\(\state^\system{U}_0 \correspond[s] \state^\system{S}_0\). Since
\simulates[\properties]{\system{S}}{\system{U}},
\[\exists \state^\system{U}_1.(\state^\system{U}_0 \Mapsto \state^\system{U}_1
\land \state^\system{U}_1 \correspond[s] \state^\system{S}_1)\]

Thus, given \(\state^\system{T}_0, \state^\system{T}_1 \in \states^\system{T},
\state^\system{U}_0 \in \states^\system{U}\) such that \(\state^\system{T}_0
\mapsto \state^\system{T}_1\) and \(\state^\system{U}_0 \correspond[s]
\state^\system{T}_0\),
\[\exists \state^\system{U}_1 \in \states^\system{U}.(\state^\system{U}_0
\Mapsto \state^\system{U}_1 \land \state^\system{U}_1 \correspond[s]
\state^\system{T}_1)\]

Hence, \simulates[\properties]{\system{T}}{\system{U}}.
\end{proof}

Next, we demonstrate step~2 (if direction):
\begin{lemma} \label{thm:als:if}
Given access control systems \system{S} and \system{T} and simulation properties
\(\properties = \propertyset\),
\(\reduction[\properties]{\system{S}}{\system{T}} \implies
\simulates[ALS]{\system{T}}{\system{S}}\).
That is, if \system{T} is at least as expressive as \system{S} with respect to
properties \properties, then \system{T} admits an ALS simulation of \system{S}.
\end{lemma}

\begin{proof}
To prove this \lcnamecref{thm:als:if}, we let \system{S} and \system{T} be
arbitrary access control systems such that
\reduction[\properties]{\system{S}}{\system{T}}, and we show that
\simulates[ALS]{\system{T}}{\system{S}}.

Since \reduction[\properties]{\system{S}}{\system{T}}, for any access control
system \system{U}, if \simulates[\properties]{\system{S}}{\system{U}}, then
\simulates[\properties]{\system{T}}{\system{U}}.

Since \system{S} can trivially simulate itself,
\simulates[\properties]{\system{S}}{\system{S}}, and thus
\simulates[\properties]{\system{T}}{\system{S}}.

Thus, given \(\state^\system{S}_0, \state^\system{S}_1 \in \states^\system{S},
\state^\system{T}_0 \in \states^\system{T}\), by forward reachability, if
\(\state^\system{S}_0 \correspond[s] \state^\system{T}_0\) and
\(\state^\system{S}_0 \mapsto \state^\system{S}_1\), then
\[\exists \state^\system{T}_1.(\state^\system{T}_0 \Mapsto \state^\system{T}_1
\land \state^\system{S}_1 \correspond[s] \state^\system{T}_1)\]

Since \propsym{SCs} and \propsym{QPa} satisfy the ALS definition of state
correspondence, this means we have satisfied the first property of the ALS
simulation.
\begin{enumerate}

\item If \system{S} can reach a given state, \system{T} can reach a
corresponding state.

\end{enumerate}

And by bidirectional reachability, given \(\state^\system{S}_0 \in
\states^\system{S}, \state^\system{T}_0, \state^\system{T}_1 \in
\states^\system{T}\), if \(\state^\system{S}_0 \correspond[s]
\state^\system{T}_0\) and \(\state^\system{T}_0 \mapsto \state^\system{T}_1\),
then
\[\exists \state^\system{S}_1.(\state^\system{S}_0 \Mapsto \state^\system{S}_1
\land \state^\system{S}_1 \correspond[s] \state^\system{T}_1)\]

And therefore, we have satisfied the second property of the ALS simulation:
\begin{enumerate} \setcounter{enumi}{1}

\item If \system{T} can reach a given state, \system{S} can reach a
corresponding state.

\end{enumerate}

These properties satisfy the definition for ALS simulation, and hence \system{T}
admits an ALS simulation of \system{S}
(\simulates[ALS]{\system{T}}{\system{S}}).
\end{proof}

Therefore, we have proved the decomposition of the ALS simulation:
\begin{theorem} \label{thm:als}
\(\decomposes{\text{ALS}}{\propertyset}\); that is, the ALS simulation
decomposes to structure correspondence, authorization preservation, and
bidirectional reachability.
\end{theorem}

\begin{proof}
By \cref{thm:als:onlyif}, if \simulates[ALS]{\system{T}}{\system{S}}, then
\reduction[\properties]{\system{S}}{\system{T}}. By \cref{thm:als:if}, if
\reduction[\properties]{\system{S}}{\system{T}}, then
\simulates[ALS]{\system{T}}{\system{S}}. Thus,
\reduction[\properties]{\system{S}}{\system{T}} if and only if
\simulates[ALS]{\system{T}}{\system{S}}, and thus the ALS simulation decomposes
to \propertyset.
\end{proof}

\inlongshort{%
All other decomposition proofs can be found in \cref{sec:proofs}.
}{%
In the interest of space, all other decomposition proofs can be found in the
technical report accompanying this paper~\cite{csf2015extended}.
}

\subsection{Results}

Now, we present the results of decomposing the simulations from the series of
previous works discussed in \cref{sec:expressiveness} into sets of simulation
properties from \cref{sec:properties}. First, a chart of our results is shown in
\cref{fig:positioning}, which states the decomposition of the SMG
simulation~\cite{Sandhu92,SG93,SG94,SM98,MS99,OSM00}, the Ganta
simulation~\cite{Ganta96}, the ALS simulation~\cite{ALS92,ALS96}, the CDM weak
and strong simulations~\cite{CDM01}, the TL state-matching
reduction~\cite{TL04,TL07}, and HMG+ parameterized expressiveness (along with
several parameterized expressiveness properties)~\cite{HMG13}. Properties are
omitted if they are not explicitly required by the simulation's definition but
are implied by other, explicit properties (e.g., CDMs decomposes to a set
including \propsym{CDi}, which also implies \propsym{CCc}).
\cref{sec:selecting:interaction} discusses which properties imply others.

\begin{figure*}
\centering
\includegraphics[width=.88\textwidth]{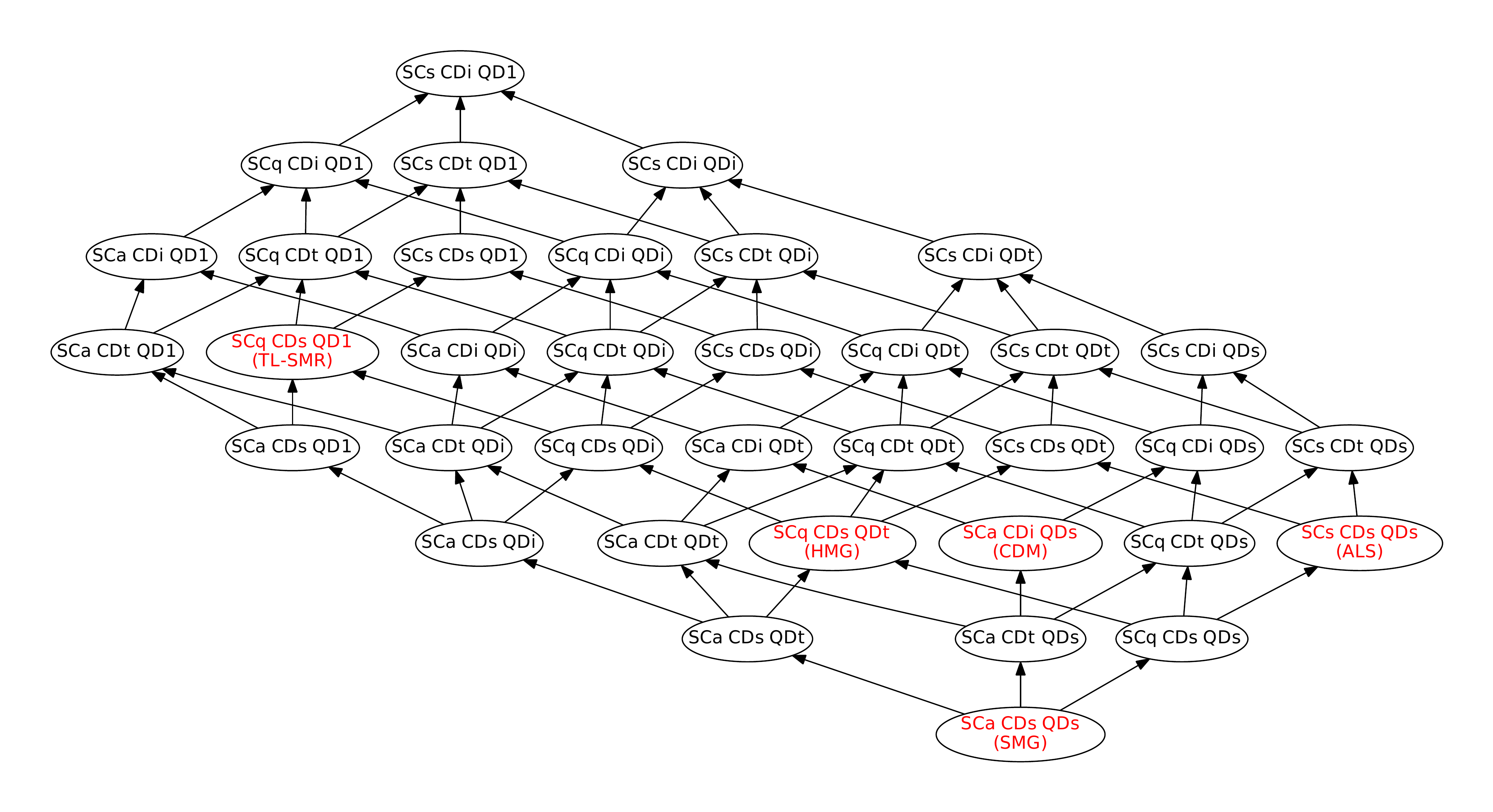}
\caption{Lattice of state correspondence, command dependence, and query
dependence with positioned surveyed simulations}
\label{fig:biglattice}
\end{figure*}

In \cref{fig:taxonomy}, we arrange this data as a taxonomy, with each split
representing a dimension, with weaker properties positioned to the left and
stronger properties to the right. We split first on the state correspondence,
which is perhaps the biggest difference among the surveyed simulations. This
separates simulations that preserve only the answers to authorization requests
(\propsym{SCa}) from those that preserve all queries (\propsym{SCq}) and those
that preserve full state structure (\propsym{SCs}). We note that the ALS
simulation is alone in its decomposition including \propsym{SCs}; all other
surveyed simulations allowed the simulating system to store information in a
different organization than the simulated system, so long as the required
queriable information (requests or queries) can be recovered. We also note that
the predominant difference between the SMG simulation and the CDM simulations is
the command dependence: in SMG, a command can be mapped completely differently
if it is to be executed in different states, while in CDM, each command must be
mapped without knowing the state in which it will be executed. The Ganta
simulation is unique in enforcing the non-contamination trace restriction. HMG+
and TL-SMR use the same state correspondence, but HMG+ enforces a more lax query
dependence and does not require bireachability. Simulations that are positioned
farther apart are the most dissimilar. Most starkly different are SMG and ALS,
positioned far left and far right, which share no simulation properties except
in dimensions in which both enforce only minimum properties, despite their
similar publication times.

In \cref{fig:partlattice}, we position the surveyed simulations within a
lattice. Higher simulations decompose to more strict properties, and an arrow
from simulation \simulation to simulation \(\simulation^\prime\) indicates that
\(\simulation^\prime\) decomposes to strictly stronger properties than
\simulation. Here we can see that the SMG simulation is strictly weakest, which
supports previous claims to this effect~\cite{TL07,Ganta96}. Several orthogonal
directions were taken in defining other simulations to enforce stronger
properties. The CDM simulations, as noted above, restrict the command
dependence. The Ganta simulation requires non-contamination and bireachability.
The TL state-matching reduction and HMG+ parameterized expressiveness consider
queries, and thus strengthen the state correspondence. The ALS simulation
enforces an even more strict state correspondence, requiring the structure of a
simulated system's state to be preserved in the simulating system.
Interestingly, we note that while all are stronger than SMG, most pairs are
incomparable due to being stronger in orthogonal ways. In particular, while
TL-SMR is considered to be a relatively strong notion of simulation, this is not
substantiated by the lattice, which shows TL-SMR to be stronger than HMG+ and
SMG, but incomparable to the CDM, ALS, and Ganta simulations.

\Cref{fig:biglattice} presents a lattice of state correspondence, command
dependence, and query dependence, with the surveyed simulations positioned
within it (in this space, the Ganta simulation is at the same point as the SMG
simulation). This \lcnamecref{fig:biglattice} makes evident the wide range of
points between existing simulations that have not been explored. In this
\lcnamecref{fig:biglattice}, we omit several dimensions for readability, namely
reachability (which further separates Ganta, ALS, and TL-SMR from SMG, CDM and
HMG+) and stuttering (which would break CDM into its weak and strong
counterparts). Perhaps the most interesting points to explore within this
lattice are those that exist between two surveyed simulations. For example,
\(\set{\propsym{SCq}, \propsym{CDs}, \propsym{QDs}}\) adds to SMG the
preservation of queries beyond requests, but stops short of HMG+ by not
restricting the query decider to consider only the theory of the state while
mapping queries. Similarly, \(\set{\propsym{SCa}, \propsym{CDt},
\propsym{QDs}}\) takes away some of SMG's freedom to inspect the state mapping
commands, but rather than go all the way to the independent command mapping of
CDM, it still allows it to inspect the state's responses to queries. We also
point out \(\set{\propsym{SCq}, \propsym{CDs}, \propsym{QDi}}\), which differs
from HMG+ by enforcing query decider independence (mapping queries cannot
consider the state or theory), but can map each simulated query to a boolean
expression over simulating queries.

%% file: selecting.tex
\section{Selecting New Sets of Properties} \label{sec:selecting}

In \cref{sec:positioning}, we positioned the simulations used in previous works
within a comparative lattice, allowing them to be formally compared for the
first time. In this \lcnamecref{sec:selecting}, we enable a second use of our
lattice of expressiveness simulation properties: crafting new notions of
expressiveness by choosing the properties that most closely correspond to the
scenario in which an access control system will be deployed. We first discuss
interactions between dimensions; this discussion should act as a warning against
choosing individual properties in isolation. We then interpret the impact each
identified dimension has on the simulation, and identify properties of a
deployment scenario that may dictate particular choices in each dimension.
Finally, we discuss the potential impact these techniques could have on future
expressiveness analysis.

\subsection{Interactions Between Dimensions} \label{sec:selecting:interaction}

We noted in \cref{sec:positioning} that some simulations decompose to sets of
properties that include \emph{implied} properties, or properties that are
redundant given the others in the set. For instance, command independence
(\propsym{CDi}) implies constant-time command mapping (\propsym{CCc}); if the
command mapping does not depend on the state, then its procedure must be
constant-time in the size of the state. Further, \propsym{CCc} implies constant
step (\propsym{CSc}), since a constant-time procedure must have constant-size
output.

An additional type of interaction is between basic properties and those
properties whose definition relies on the basic properties in the abstract. For
example, the definition of forward reachability (\propsym{R\reach}) refers to
sequences of commands output by \(\mapping^\commands\), the length of which may
be limited by command mapping stuttering (\propsym{CS}). Further, the
definitions of both reachability properties (\propsym{R}) and trace structure
properties (\propsym{CT}) refer to corresponding states. Here, the details of
what makes states correspond is left to the state correspondence structure
(\propsym{SC}).

These dependencies show that the proof of a property in one dimension may rely
on the properties chosen in another. Thus, e.g., changing to a stronger state
correspondence requires re-proving a simulation's results for reachability and
trace structure, since these are dependent on state correspondence.

Several property dimensions are defined over the size of the simulated state:
command mapping complexity (\propsym{CC}), command mapping stuttering
(\propsym{CS}), and query decider time-complexity (\propsym{QC}). Thus, these
dimensions can be altered with respect to the original, simulated state by the
state storage size (\propsym{SS}). For example, enforcing polynomial storage
(\propsym{SSp}) and linear-time command mapping (\propsym{CCl}) will guarantee a
command mapping that is linear-time with respect to the \emph{simulating state},
which is a polynomial expansion over the original \emph{simulated state}.

\subsection{Interpreting the Dimensions} \label{sec:selecting:interpret}

We now discuss the practical impacts each identified dimensions, and what types
of environments may cause one to prefer a particular property in these
dimensions over others.

\bprop{SC}{State correspondence structure} allows one to change what needs to be
preserved about the state during a simulation. If the deployment scenario in
question assumes only that the simulation allows the proper authorization
requests, \propsym{SCa} should suffice. For scenarios that require the access
control system to support (and provide correct answers to) additional queries
such as, \emph{``Is user \(u\) a member of role \(r\)?''}, \propsym{SCq} is more
appropriate. Finally, in scenarios that make use of additional code that has
access to (and assumes a particular arrangement of) the access control system's
internal data structures, \propsym{SCs} is the best choice.

\bprop{SS}{State storage} limits the size of the simulated state with respect to
the original state (i.e., the state of the system being simulated). This can be
restricted for several reasons. The most obvious is storage space: if trusted
storage for representing access control state is limited, we may restrict the
simulation from mapping states in a way that increases storage by more than a
linear factor (\propsym{SSl}) or a polynomial factor (\propsym{SSp}). However,
the more interesting reason comes from an interaction described in
\cref{sec:selecting:interaction}. Since other dimensions place restrictions
(e.g., on the number of commands executed) based on the size of the simulating
state, we may restrict the state expansion to linear (\propsym{SSl}) in order,
e.g., to restrict the command mapping procedure to be linear-time in the size of
the \emph{original, simulated state}. If state storage is polynomial
(\propsym{SSp}), then even if we enforce a command mapping that is linear in the
simulating state (\propsym{CCl}), this only restricts it to being
polynomial-time with respect to the simulated state. Thus, even when trusted
storage space is unbounded in the deployment scenario, one may desire to limit
state size to limit later iteration over this state.

\bprop{CD}{Command mapping dependence} allows one to require that the command
mapping be computable without full knowledge and inspection of the state in
which a command will be executed in. Independent command (\propsym{CDi})
requires that each command is mapped independent of the state, and is useful in
deployment scenarios in which the agent calculating the simulating commands is
completely unprivileged, and cannot inspect the state. It is also useful when
commands must be precompiled, thus adding no computation at runtime beyond that
of the simulating commands themselves. Theory-dependent command mapping
(\propsym{CDt}) allows the command mapping to inspect the \emph{theory} of the
state (i.e., the answers to all queries). This property is useful in deployment
scenarios in which the simulation agent is no more privileged than normal
users---calculating the mapped commands requires only information available by
asking queries. Finally, state-dependent command mapping (\propsym{CDs}) allows
the command mapping to arbitrarily inspect the state, requiring a powerful
simulation agent.

\bprop{CC}{Command mapping complexity} restricts the time-complexity of the
command mapping with respect to the size of the simulating state. Constant
command mapping (\propsym{CCc}) can restrict the command mapping from taking any
longer for larger states, and is thus appropriate when states can be large but
mapping commands must always remain fast. Linear command mapping (\propsym{CCl})
prevents expensive nested loops over access control state as well as operations
such as sorting, while still allowing more processing for larger states.

\bprop{CS}{Command stuttering} restricts the number of simulating commands
executed for each simulated command. Lock-step (\propsym{CS1}) simulations must
execute no more than one simulating command per simulated command, and thus
ensure there is no intermediate state exposed to users. In deployment scenarios
without the ability to force atomic execution of a sequence of commands (or
without built-in data structure locking), this property is crucial to preventing
the inspection of intermediate (potentially inconsistent) states. Constant step
(\propsym{CSc}) simulations are allowed a constant number of commands for each
simulated command, and are thus appropriate when the state can grow to be large
but the deployment scenario requires that the number of steps for any simulated
action remain bounded (e.g., to prevent starvation due to locked structures).

\bprop{CT}{Trace structure} properties restrict the path that the simulating
system can take during the simulation of a single command. Semantic lock-step
(\propsym{CT1}, depicted in \cref{fig:semanticlock}) provides the benefits of a
lock-step simulation in a slightly relaxed way: a ``setup'' phase prepares for
the transition by changing only internal data (i.e., while remaining equivalent
to the start state), then the transition occurs to a state equivalent to the end
state, and then the ``cleanup'' phase cleans up any unnecessary leftover data
(again, while remaining equivalent to the same end state). This is particularly
useful when lock-step is too strict, but the deployment scenario is sensitive to
the exposure of intermediate states (since, in \propsym{CT1}, no states are
exposed except those equivalent to the start and end states). Query monotonicity
(\propsym{CTq}) ensures that no query changes its truth value except those that
are required to change between the start and end state. This allows multiple
steps, but ensures that intermediate states, while not corresponding with the
start or end state, never answer any query in a way that neither the start nor
end state would. This is useful in scenarios where intermediate states are
undesirable, but users are not expected to execute more than a single query
between ``valid'' states (and will thus never detect the inconsistency). Access
monotonicity (\propsym{CTa}) is similar, but applies only to authorization
requests, and is useful in scenarios where inconsistent states are not a danger
as long as they do not wrongly allow or forbid a request. Finally,
non-contamination (\propsym{CTs}) ensures that no two accesses are allowed in an
intermediate state that are not \emph{both} allowed in either the start or end
state. Thus, the simulating system is restricted not only from allowing accesses
forbidden in the simulated system, but also combinations of individually-allowed
accesses that are never combined in the simulated system. This restriction is
particularly useful in environments with operations that ``swap'' accesses from
one subject or object to another, or where separation of privilege is utilized.

\bprop{CA}{Actor preservation} restricts which users can be invoked to simulate
commands. Self-execution (\propsym{CA\top}) requires each simulating command be
executed by the same user as the original, simulated command. This allows the
simulating agent to be completely unprivileged, mapping commands as a service to
the user, but without executing them with any privilege beyond the user's own.
Administration-preservation (\propsym{CAa}) requires any non-administrative
simulated command be mapped to a sequence of non-administrative commands (i.e.,
a command that does not invoke administrative privileges cannot be simulated by
an administrative command). This corresponds to scenarios in which users will be
expected to operate largely without administrative intervention. No restriction
in this dimension means that the command mapping can return commands to be
executed by any other user. This is most appropriate when the simulating agent
is trusted to execute administrative actions on behalf of untrusted users, or
when the commands returned can then be delegated to other users to be approved
and executed.

Finally, \bprop{R}{reachability} specifies whether the simulating system should
be restricted from entering a state that does not correspond to a simulated
state. If the simulation agent is users' only interface to the deployed access
control system, forward reachability (\propsym{R\reach}) is sufficient. However,
if users can access the simulating system's native commands, bireachability
(\propsym{R\bireach}) ensures that the system cannot transition to a state that
is inconsistent with the simulated system.

\subsection{Studying Canonical Usages} \label{sec:selecting:canonical}

\begin{figure}
\centering
\includegraphics[width=0.72\columnwidth]{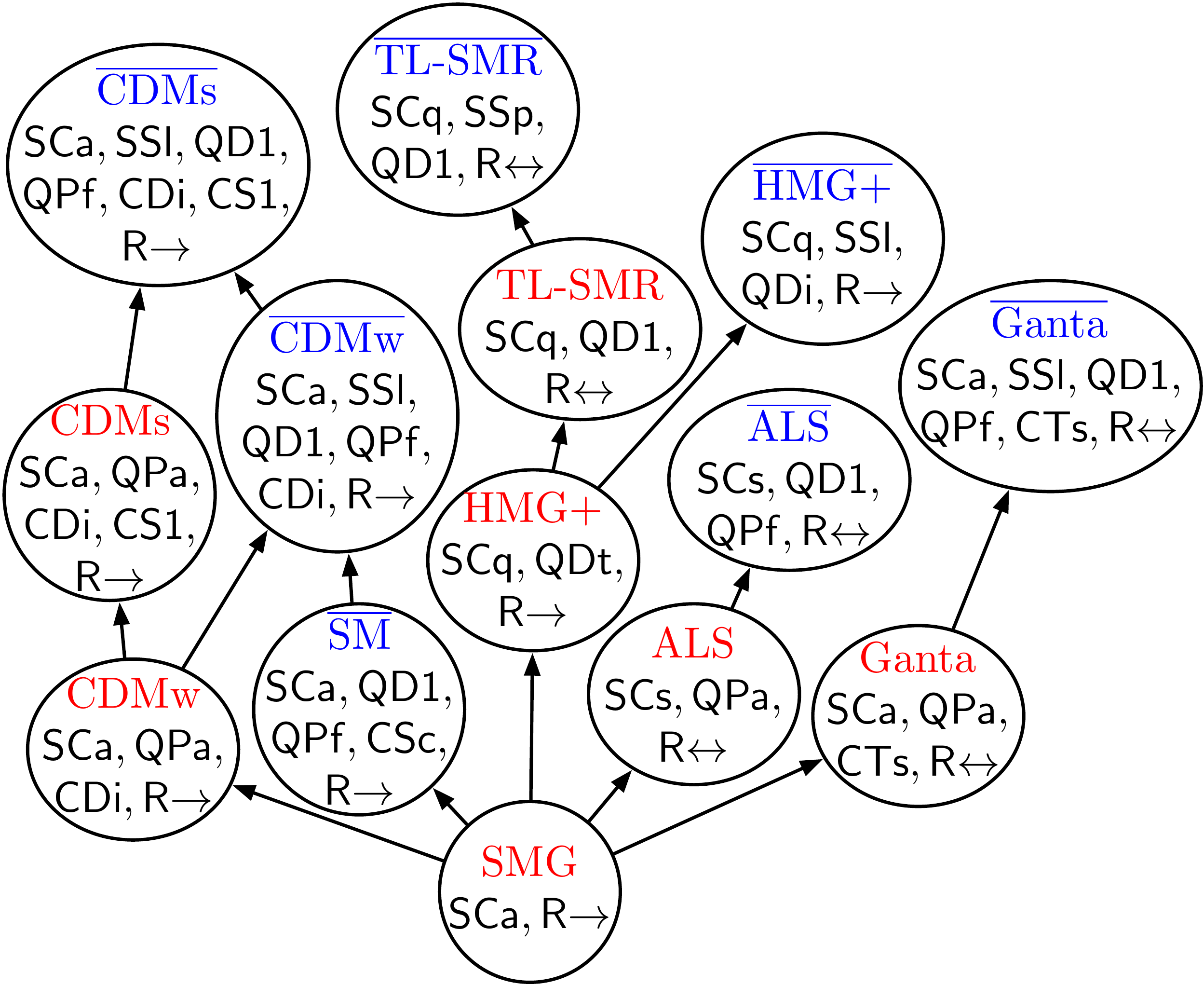}
\caption{Partial lattice of canonical usage}
\label{fig:canonlattice}
\end{figure}

Next, we use the above interpretation of our expressiveness simulation
properties to guide a discussion about how each of the notions of simulation
that we studied in \cref{sec:positioning} is used by its creators. In many
cases, the definition for a particular notion of simulation is underconstrained,
and the simulations written within the framework actually satisfy stronger
properties than the defined lower bound. We refer to the set of properties that
the authors seem to intend for a simulation to uphold as its \emph{canonical
usage}. In the case of Sandhu's simulation, the author recognizes that the given
constructions are stronger than the definition, noting that formalizing the
definition of the stronger simulation is beyond the scope of the
work~\cite{Sandhu92}. Here, we make conjectures regarding the decomposition of
the canonical usage of these simulations. A lattice view of these conjectures is
shown in \cref{fig:canonlattice}, where \canonical{\simulation} indicates the
canonical usage of simulation type \simulation. For example, \canonical{SM}
refers to the form of the SMG simulation used in \cite{SM98,MS99}.

It is interesting to note that the relationships between notions of simulation
are not necessarily preserved in the canonical usage. While SMG by definition is
the weakest simulation, the canonical usage \canonical{SM} is incomparable to
any simulation's definition and positioned strictly weaker than the canonical
usage of the CDM simulations. While, by definition, the TL state-matching
reduction is more strict than HMG+ parameterized expressiveness, their canonical
usages are incomparable due to \canonical{TL-SMR} enforcing bireachability
(\propsym{R\bireach}) and using polynomial state size (\propsym{SSp}), compared
to \canonical{HMG+} enforcing forward reachability (\propsym{R\reach}) and using
linear state size (\propsym{SSl}). Finally, we note that all of CDMs, CDMw, SMG,
and ALS simulations are canonically used in such a way that enforces full query
preservation (\propsym{QPf}); that is, all of the constructed mappings of these
types use the identity mapping for all supported queries, despite the fact that
none of them specifically require this by definition. This trend of a notion of
simulation's usage being consistently more strict than its definition reveals
the difficulty in fully specifying the set of properties that a notion of
expressiveness simulation is intended to enforce. The discussion in this
section, aimed at helping analysts choose a reasonable set of properties for a
deployment, can also help ensure that newer notions of simulation are fully
specified, and best match their intended usages.

%% file: conclusion.tex
\section{Conclusion and Future Work} \label{sec:conclusion}

In this paper, we organize the existing knowledge of expressiveness simulations
by formalizing a granular, property-based representation, proposing a wide range
of dimensions of simulation properties, and positioning influential notions of
expressiveness simulation from the literature within the lattice of these
properties. In doing so, we provide the first systematic comparison of existing
simulations that were not previously known to be directly comparable, showing
how these notions of expressiveness simulation relate to one another.

Looking away from existing notions of simulation and rather \emph{between} them,
this work allows us to explore an organized space of simulations to identify
areas to explore in future research. For instance, knowing expressiveness
results derived using the SMG and ALS simulations, which of these hold true for
notions of simulation ``between'' the two existing notions? What results can be
shown for a simulation decomposing to the union of the properties of two
existing notions? How far up the lattice do all systems become incomparable?
These questions can only be explored thanks to the systematic means of
simulation decomposition.

Finally, understanding the systems implications of various simulation properties
will enable analysts to select the notion of access control expressiveness that
corresponds most closely to the scenario in which they plan to deploy the target
access control system(s). Thus, we make inroads toward bringing expressiveness
analysis techniques out of the strictly formal realm, and repurpose these
techniques to help select the most suitable access control system for a given
application.

A question to be explored in future work is the identification of the set of
analysis questions that a particular set of simulation properties preserve. For
example, Tripunitara and Li showed that the state-matching reduction preserves
\emph{compositional security analysis instances}: the set of questions
containing a single quantifier (\(\exists\) or \(\forall\)), a propositional
formula over queries \(\varphi\), and a start state \state~\cite{TL07}.
Semantically, the question asks whether \(\varphi\) it is \{ever, always\} true
in states reachable from \state. If \system{T} admits a state-matching reduction
of \system{S}, then all compositional security analysis instances have the same
value in \system{S} and \system{T}. Identifying the types of analysis questions
preserved by other notions of simulation would allow us even greater
understanding of the practical and theoretical impacts of simulation property
choices.

%% file: implementations.bbl

%% file: queryproperties.tex
\section{Query decider properties} \label[appendix]{sec:queryproperties}

In this \lcnamecref{sec:queryproperties}, we present the dimensions in which the
query decider can be restricted. Recall (\cref{def:implementable}) that the
query decider is a function \(\mapping_\queries = \queries^\system{S} \times
\states^\system{T} \to \set{\true, \false}\) that assigns a truth value to each
query in system \system{S}, in each state in system \system{T}. As with the
command mapping, we can restrict what properties of the state the query decider
can observe, but in general it sees the whole state. Thus, it allows us to
answer \system{S} queries in an active simulation using \system{T}. We now
discuss the ways in which we can restrict this decider.

\dimension{QD}{Query decider dependence}

While \cref{def:implementable} maps each \system{S} query and \system{T} state
to a truth value, previous works use more strict query deciders, ranging from
mapping each \system{S} query \query to a \system{T} query \(\query^\prime\) and
returning \(\state^\system{T} \entails \query^\prime\)~\cite{TL07}, to mapping
each \system{S} query and \system{T} \emph{theory} to a truth value~\cite{HMG13}
(calculating the truth value by observing only the queriable portions of the
\system{T} state). \emph{Query decider dependence}, similar to command mapping
dependence, restricts the information that the query decider can consider about
a \system{T} state when deciding the truth value of an \system{S} query in that
state.

\begin{sproperty}[\prop{QD1}{Independent, unitary-range query decider}]
\(\exists \mapping^\prime: \queries^\system{S} \to
\queries^\system{T}.(\mapping_\queries\prm{\query, \state} \equiv \state
\entails \mapping^\prime\prm{\query})\)
\end{sproperty}

\begin{sproperty}[\prop{QDi}{Independent query decider}]
\(\exists \mapping^\prime: \queries^\system{S} \to
\boolexp{\queries^\system{T}}.(\mapping_\queries\prm{\query, \state} \equiv
\state \entails \mapping^\prime\prm{\query})\)
\end{sproperty}

\begin{sproperty}[\prop{QDt}{Theory-dependent query decider}]
\(\exists \mapping^\prime: \queries^\system{S} \times \theory{\system{T}} \to
\set{\true, \false}.\mapping_\queries\prm{\query, \state} \equiv
\mapping^\prime\prm{\query, \theory{\state}}\)
\end{sproperty}

\begin{property}[\prop{QDs}{State-dependent query decider}]
\nullproptext
\end{property}

First, \emph{independent, unitary-range query} simulations first map each
\system{S} query \(\query^\system{S}\) to a single \system{T} query
\(\query^\system{T}\), then return the truth value of asking this \system{T}
query in \(\state^\system{T}\) (i.e., returns \(\state^\system{T} \entails
\query^\system{T}\)). This type of simulation was used in \cite{TL07}.
\emph{Independent query} simulations, similarly, map \(\query^\system{S}\)
independent of the state, but in this case can map it to an element of
\(\boolexp{\queries^\system{T}}\), a boolean expression over
\(\queries^\system{T}\), rather than a single query. This allows the simulation,
for example, to map the \system{S} query ``Is user \(u\) a member of role
\(\request\)'' to \system{T} query sentence ``Does user \(u\) exist and does
user \(u\) have attribute \(\text{role}\mathord:r\)?''

\emph{Theory-dependent query} allows the truth value to be determined based only
on the \emph{theory} of the \system{T} state, or the values of all \system{T}
queries in the state. This allows the query decider to inspect the values of
(potentially) all \system{T} queries, but does not allow the decider to consider
any features of the state that cannot be queried using \(\queries^\system{T}\).
This version of the query decider is used in \cite{HMG13}. Finally, we refer to
the general case as \emph{state-dependent query}, since in this case the query
decider can arbitrarily inspect the \system{T} state before returning a truth
value for an \system{S} query, rather than being restricted only to those
elements of the state which are observable by asking queries in
\(\queries^\system{T}\).

The biggest impact a selection in \propsym{QD} has is to limit the privilege of
the simulating agent: under \propsym{QD1} and \propsym{QDi}, the simulating
agent need know nothing about the current state of the system to map queries,
and simulated queries are mapped statically to dynamic queries. Under
\propsym{QDt}, the simulating agent must be privileged to view the values of all
queries at runtime, while \propsym{QDs} assumes the ability to arbitrarily
inspect the state at runtime (even the portions of state that are not
queriable).

\dimension{QC}{Query decider complexity}

As with the command mapping, due to potential resource constraints, we may
enforce limits on the runtime complexity of the query decider routine, with
respect to the size of the state it is executed in. As with commands, this is
not applicable for independent query deciders.

\begin{property}[\prop{QCc}{Constant query decider}]
\(\forall \query \in \queries^\system{S}\), the algorithm for
\(\mapping_\query\prm{\state} = \mapping_\queries\prm{\query, \state}\) has time
complexity \(T\prm{n} \in \bigO{1}\)
\end{property}

\begin{property}[\prop{QC\infty}{Unbounded query decider}]
\nullproptext
\end{property}

For theory-dependent and state-dependent query deciders, we can limit the
complexity of the procedure---here, we consider only the constant-time
restriction explicitly, though other restrictions may be useful in some cases.

\dimension{QP}{Query preservation}

Query preservation is a property dimension that indicates which queries need to
stay the same as they are mapped from system \system{S} to system \system{T}. A
particular application may require any given set of queries to be preserved;
here, we present several generic examples.

\begin{sproperty}[\prop{QPf}{Complete query preservation}]
\(\forall \query \in \queries^\system{S}, \state \in \states^\system{T} :
\mapping_\queries\prm{\query, \state} \equiv \state \entails \query\)
\end{sproperty}

\begin{sproperty}[\prop{QPa}{Authorization preservation}]
\(\forall \request \in \requests^\system{S}, \state \in \states^\system{T} :
\mapping_\queries\prm{\request, \state} \equiv \state \entails \request\)
\end{sproperty}

\begin{sproperty}[\prop{QPw}{Weak authorization preservation}]
\(\exists f : \requests^\system{S} \to \requests^\system{T}\) such that the
following two conditions hold:
\begin{itemize}

\item \(\forall \request \in \requests^\system{S}, \state \in
\states^\system{T}.(\mapping_\queries\prm{\request, \state} = \true \implies
\state \entails f\prm{\request})\)

\item \(\forall \request \in \requests^\system{T}, \state \in
\states^\system{T}.(\state \entails \request \implies \exists
\request^\system{S}.(\mapping_\queries\prm{\request^\system{S}, \state} = \true
\land f\prm{\request^\system{S}} = \request))\)

\end{itemize}
\end{sproperty}

The most common property in this dimension is \emph{authorization preservation},
which roughly enforces that the query decider maps each \system{S} request to
the value of the identical request in the \system{T} state. This requires
\system{T} to accept at least the same authorization requests as \system{S}, and
can be seen as ensuring that \system{T} is using its model ``as intended''
(i.e., forcing it to answer simulated requests as it would its own native
requests). Complete query preservation restricts the query decider in the same
way, but for all supported queries.

Authorization preservation was formalized in \cite{HMG13} (as AC-preservation),
but has been used implicitly in other simulations (e.g., \cite{CDM01}) that do
not include any mapping from \system{S} requests to \system{T} requests (i.e.,
assume the identity mapping). For formalisms that do not consider queries other
than requests, authorization preservation and complete query preservation are
equivalent.

A related property is \emph{weak authorization preservation}, defined in
\cite{codaspy} (as weak AC-preservation). This property is a weakened version of
authorization preservation: its intentions are similar, but the weak form can be
used even when \system{S} and \system{T} do not support the exact same requests
(i.e., simulating a system with requests of the form, ``Does user \(u\) have
access to permission \(p\)?'' in a system with requests of the form, ``Does
subject \(s\) have access \({read}\) to object \(o\)?''). Weak authorization
preservation allows a request transformation function, which maps \system{S}
requests to \system{T} requests. The definition of this property ensures that
each \system{S} request is mapped into \system{T}, and each \system{T} request
that is granted represents some \system{S} request.

\subsection{Interactions Between Dimensions}
\label{sec:queryproperties:interaction}

Independence query decider (\propsym{QDi}) implies constant query decider
(\propsym{QCc}), since a query decider that does not depend on the state must be
constant-time with respect to the state size. Full query preservation
(\propsym{QPf}) implies constant query decider (\propsym{QCc}), since the
identity mapping is a constant-time procedure, and implies unitary-range query
decider (\propsym{QD1}), since the identity mapping always outputs only one
query.

\subsection{Interpreting the Dimensions} \label{sec:queryproperties:interpret}

\bprop{QD}{Query dependence}, similar to command dependence, can restrict the
query decider from using the full knowledge of the state in question when
mapping queries to truth values. Independent query decider (\propsym{QD1} and
\propsym{QDi}) map each simulated query to its simulating queries, and its truth
value is then determined by checking the values of these mapped queries in the
simulating state. This is useful, as with commands, for precompilation, and
restricts the simulating agent's role to be akin to a Karp reduction (i.e., it
can only return the value of its simulating queries with no
modifications)~\cite[p.~60]{karp}. Theory-dependent query decider
(\propsym{QDt}) allows the inspection of the full theory of the state, and thus
arbitrary computation over the values of all queries in the simulating system,
corresponding to a query decider that is only privileged enough to ask queries,
but not inspect state. Finally, state-dependent query decider (\propsym{QDs})
allows arbitrary inspection of the state, and thus may answer simulated queries
using state that is usually unqueriable in the simulating system; this requires
a powerful simulation agent that is trusted to view the full access control data
structures.

\bprop{QC}{Query complexity} restricts the time-complexity of the query decider
with respect to the size of the simulating state, and is thus chosen for reasons
analogous to command complexity (\propsym{CC}).

\bprop{QP}{Query preservation} restricts certain queries to be mapped by a sort
of ``identity function.'' That is, certain simulated queries are mapped \true if
and only if the query is also present in the simulating system, and the query
returns \true in the current simulating state. This dimension of restrictions is
strict in that it requires the simulating system to support all of the included
queries of the simulated system, but ensures that the simulating system is used
as per normal. If the deployment scenario is unable to cope with reduced
throughput caused by the query decider during simulation, \propsym{QPf} also
ensures that we can pipe simulated queries directly to the simulating system.
Particularly useful is authorization preservation (\propsym{QPa}) when using a
simulating system based on a model with formally-proven properties; Since all
authorization requests must be mapped by the identity, the simulating system
allows a simulated request exactly when it would natively allow the same
request.

%% file: proofs.tex
\section{Decomposition Proofs} \label[appendix]{sec:proofs}


\input{proofs/tlsmr}

\input{proofs/hmg}

\input{proofs/hmgetc}
\input{proofs/smg}

\input{proofs/ganta}

\input{proofs/cdmw}

\input{proofs/cdms}

%% file: proofs/tlsmr.tex
\subsection{TL State-Matching Reduction}

\renewcommand{\propertyset}{\ensuremath{\set{\propsym{SCq}, \propsym{QD1}, \propsym{R\bireach}}}\xspace}

For the purposes of this proof, let the set of simulation properties
\(\properties = \propertyset\).
\begin{lemma} \label{thm:tl:onlyif}
Given access control systems \system{S}, \system{T}, and \system{U},
\[\simulates[TL-SMR]{\system{T}}{\system{S}} \land
\simulates[\properties]{\system{S}}{\system{U}} \implies
\simulates[\properties]{\system{T}}{\system{U}}\]
That is, if \system{T} admits a state-matching reduction of \system{S}, and
\system{S} admits a simulation of \system{U} with properties \propertyset, then
\system{T} admits a simulation of \system{U} with properties \propertyset.
\end{lemma}

\begin{proof}
Let \system{S}, \system{T}, and \system{U} be arbitrary access control systems
such that \simulates[TL-SMR]{\system{T}}{\system{S}} and
\simulates[\properties]{\system{S}}{\system{U}}. To prove \cref{thm:tl:onlyif},
we must then show that \simulates[\properties]{\system{T}}{\system{U}}.


Since \simulates[\properties]{\system{S}}{\system{U}}, by \propsym{QD1},
\[\exists \mapping^{QD1}: \queries^\system{U} \to
\queries^\system{S}.(\mapping_\queries\prm{\query^\system{U}, \state^\system{S}}
\equiv \state^\system{S} \entails \mapping^{QD1}\prm{\query^\system{U}})\]

Since \simulates[TL-SMR]{\system{T}}{\system{S}}, the state-matching reduction
provides a mapping from \(\queries^\system{S}\) to \(\queries^\system{T}\). Call
this mapping \(\mapping^{SMR}\).

Thus, let \(\mapping^\prime: \queries^\system{U} \to \queries^\system{T} =
\mapping^{SMR} \cdot \mapping^{QD1}\), and say
\(\mapping_\queries\prm{\query^\system{U}, \state^\system{T}} \equiv
\state^\system{T} \entails \mapping^\prime\prm{\query^\system{U}})\). This forms
a query decider that satisfies \propsym{QD1}.


Choose an arbitrary state \(\state^\system{U}_0 \in \states^\system{U}\) and
command \(\command^\system{U} \in \commands^\system{U}\), and let
\(\trans\prm{\state^\system{U}_0, \command^\system{U}} = \state^\system{U}_1\).
Let \(\state^\system{S}_0 \in \states^\system{S}\) such that
\(\state^\system{U}_0 \correspond[q] \state^\system{S}_0\). Since
\simulates[\properties]{\system{S}}{\system{U}},
\[\exists \state^\system{S}_1 \in
\states^\system{S}.(terminal\prm{\state^\system{S}_0,
\mapping_\commands\prm{\command^\system{U}, \state^\system{S}_0}} =
\state^\system{S}_1 \land \state^\system{U}_1 \correspond[q]
\state^\system{S}_1)\]

Let \(\state^\system{T}_0 \in \states^\system{T}\) such that
\(\state^\system{S}_0 \correspond[q] \state^\system{T}_0\). Since
\simulates[TL-SMR]{\system{T}}{\system{S}},
\[\exists \state^\system{T}_1 \in \states^\system{T}.(\state^\system{T}_0
\Mapsto \state^\system{T}_1 \land \state^\system{S}_1 \correspond[q]
\state^\system{T}_1)\]
Thus, there exists a sequence of \system{T} commands \(\commands^\system{T}_0\)
such that \(terminal\prm{\state^\system{T}_0, \commands^\system{T}_0} =
\state^\system{T}_1\). Define \(\mapping_\commands: \commands^\system{U} \times
\states^\system{T} \to \prm{\commands^\system{T}}^*\) such that it returns
\(\commands^\system{T}_0\) for \(\state^\system{T}_0, \command^\system{U}\).
This is formed by concatenating a sequence of sequences of commands: for each
command \(\command^\system{S}_i\) that \system{S} needs to execute to simulate
\(\command^\system{U}\), concatenate the commands that \system{T} needs to
execute to simulate \(\command^\system{S}_i\).

Then, given \(\state^\system{U}_0, \state^\system{U}_1 \in \states^\system{U},
\state^\system{T}_0 \in \states^\system{T}, \command^\system{U} \in
\commands^\system{U}\) such that \(\trans\prm{\state^\system{U}_0,
\command^\system{U}} = \state^\system{U}_1\), and \(\state^\system{U}_0
\correspond[q] \state^\system{T}_0\),
\[\exists \state^\system{T}_1 \in
\states^\system{T}.(terminal\prm{\state^\system{T}_0,
\mapping_\commands\prm{\command, \state^\system{T}_0}} = \state^\system{T}_1
\land \state^\system{S}_1 \correspond[q] \state^\system{T}_1)\]

Hence, \simulates[\set{\propsym{SCq}, \propsym{QD1},
\propsym{R\reach}}]{\system{T}}{\system{U}}. Next we show \propsym{R\bireach}.


Choose some arbitrary states \(\state^\system{T}_0, \state^\system{T}_1 \in
\states^\system{T}\) such that \(\state^\system{T}_0 \mapsto
\state^\system{T}_1\). Let \(\state^\system{S}_0 \in \states^\system{S}\) such
that \(\state^\system{S}_0 \correspond[q] \state^\system{T}_0\). Since
\simulates[TL-SMR]{\system{T}}{\system{S}},
\[\exists \state^\system{S}_1.(\state^\system{S}_0 \Mapsto \state^\system{S}_1
\land \state^\system{S}_1 \correspond[q] \state^\system{T}_1)\]

Let \(\state^\system{U}_0 \in \states^\system{U}\) such that
\(\state^\system{U}_0 \correspond[q] \state^\system{S}_0\). Since
\simulates[\properties]{\system{S}}{\system{U}},
\[\exists \state^\system{U}_1.(\state^\system{U}_0 \Mapsto \state^\system{U}_1
\land \state^\system{U}_1 \correspond[q] \state^\system{S}_1)\]

Thus, given \(\state^\system{T}_0, \state^\system{T}_1 \in \states^\system{T},
\state^\system{U}_0 \in \states^\system{U}\) such that \(\state^\system{T}_0
\mapsto \state^\system{T}_1\) and \(\state^\system{U}_0 \correspond[q]
\state^\system{T}_0\),
\[\exists \state^\system{U}_1 \in \states^\system{U}.(\state^\system{U}_0
\Mapsto \state^\system{U}_1 \land \state^\system{U}_1 \correspond[q]
\state^\system{T}_1)\]

Hence, \simulates[\properties]{\system{T}}{\system{U}}.
\end{proof}

\begin{lemma} \label{thm:tl:if}
Given access control systems \system{S} and \system{T} and simulation properties
\(\properties = \propertyset\),
\(\reduction[\properties]{\system{S}}{\system{T}} \implies
\simulates[TL-SMR]{\system{T}}{\system{S}}\).
That is, if \system{T} is at least as expressive as \system{S} with respect to
properties \properties, then \system{T} admits a state-matching reduction of
\system{S}.
\end{lemma}

\begin{proof}
Let \system{S} and \system{T} be arbitrary access control systems such that
\reduction[\properties]{\system{S}}{\system{T}}. Since
\reduction[\properties]{\system{S}}{\system{T}}, for any access control system
\system{U}, if \simulates[\properties]{\system{S}}{\system{U}}, then
\simulates[\properties]{\system{T}}{\system{U}}.

Since \system{S} can trivially simulate itself,
\simulates[\properties]{\system{S}}{\system{S}}, and thus
\simulates[\properties]{\system{T}}{\system{S}}.

By \propsym{QD1},
\(\exists \mapping^\prime: \queries^\system{S} \to
\queries^\system{T}.(\mapping_\queries\prm{\query^\system{S}, \state^\system{T}}
\equiv \state^\system{T} \entails \mapping^\prime\prm{\query^\system{S}})\)
Thus, \(\mapping^\prime\) satisfies the format of the TL-SMR query mapping
(i.e., \(\mapping_\queries : \queries^\system{S} \to \queries^\system{T}\)).

Then, given \(\state^\system{S}_0, \state^\system{S}_1 \in \states^\system{S},
\state^\system{T}_0 \in \states^\system{T}\), by \propsym{R\reach}, if
\(\state^\system{S}_0 \correspond[q] \state^\system{T}_0\) and
\(\state^\system{S}_0 \mapsto \state^\system{S}_1\), then
\[\exists \state^\system{T}_1.(\state^\system{T}_0 \Mapsto \state^\system{T}_1
\land \state^\system{S}_1 \correspond[q] \state^\system{T}_1)\]

Since \propsym{SCq} satisfies the TL-SMR definition of state correspondence,
this means we have satisfied the first property of the state-matching reduction.
\begin{enumerate}

\item For every state \(\state^\system{S}_1\) in system \system{S} such that
\(\state^\system{S}_0 \Mapsto \state^\system{S}_1\), there exists a state
\(\state^\system{T}_1\) such that \(\state^\system{T}_0 \Mapsto
\state^\system{T}_1\) and \(\state^\system{S}_1\) and \(\state^\system{T}_1\)
are equivalent under \mapping.

\end{enumerate}

And by bidirectional reachability, given \(\state^\system{S}_0 \in
\states^\system{S}, \state^\system{T}_0, \state^\system{T}_1 \in
\states^\system{T}\), if \(\state^\system{S}_0 \correspond[q]
\state^\system{T}_0\) and \(\state^\system{T}_0 \mapsto \state^\system{T}_1\),
then
\[\exists \state^\system{S}_1.(\state^\system{S}_0 \Mapsto \state^\system{S}_1
\land \state^\system{S}_1 \correspond[q] \state^\system{T}_1)\]

And therefore, we have satisfied the second property of the state-matching
reduction:
\begin{enumerate} \setcounter{enumi}{1}

\item For every state \(\state^\system{T}_1\) in system \system{T} such that
\(\state^\system{T}_0 \Mapsto \state^\system{T}_1\), there exists a state
\(\state^\system{S}_1\) such that \(\state^\system{S}_0 \Mapsto
\state^\system{S}_1\) and \(\state^\system{T}_1\) and \(\state^\system{S}_1\)
are equivalent under \mapping.

\end{enumerate}

These properties satisfy the definition for a state-matching reduction, and
hence \system{T} admits a state-matching reduction of \system{S}
(\simulates[TL-SMR]{\system{T}}{\system{S}}).
\end{proof}

\begin{theorem} \label{thm:tl}
\(\decomposes{\text{TL-SMR}}{\propertyset}\); that is, the TL state-matching
reduction decomposes to query correspondence; independent, unitary-range query;
and bidirectional reachability.
\end{theorem}

\begin{proof}
By \cref{thm:tl:onlyif}, if \simulates[TL-SMR]{\system{T}}{\system{S}}, then
\reduction[\properties]{\system{S}}{\system{T}}. By \cref{thm:tl:if}, if
\reduction[\properties]{\system{S}}{\system{T}}, then
\simulates[TL-SMR]{\system{T}}{\system{S}}. Thus,
\reduction[\properties]{\system{S}}{\system{T}} if and only if
\simulates[TL-SMR]{\system{T}}{\system{S}}, and thus the state-matching
reduction decomposes to \propertyset.
\end{proof}

%% file: proofs/hmg.tex
\subsection{HMG+ Parameterized Expressiveness Simulation}

\renewcommand{\propertyset}{\ensuremath{\set{\propsym{SCq}, \propsym{QDt}, \propsym{R\reach}}}\xspace}

For the purposes of this proof, let the set of simulation properties
\(\properties = \propertyset\).
\begin{lemma} \label{thm:hmg:onlyif}
Given access control systems \system{S}, \system{T}, and \system{U},
\[\simulates[HMG+]{\system{T}}{\system{S}} \land
\simulates[\properties]{\system{S}}{\system{U}} \implies
\simulates[\properties]{\system{T}}{\system{U}}\]
That is, if \system{T} admits an HMG+ simulation of \system{S}, and \system{S}
admits a simulation of \system{U} with properties \propertyset, then \system{T}
admits a simulation of \system{U} with properties \propertyset.
\end{lemma}

\begin{proof}
Let \system{S}, \system{T}, and \system{U} be arbitrary access control systems
such that \simulates[HMG+]{\system{T}}{\system{S}} and
\simulates[\properties]{\system{S}}{\system{U}}. To prove \cref{thm:hmg:onlyif},
we must then show that \simulates[\properties]{\system{T}}{\system{U}}.


Since \simulates[\properties]{\system{S}}{\system{U}}, by \propsym{QDt},
\[\exists \mapping^\prime: \queries^\system{U} \times \theory{\system{S}} \to
\set{\true, \false}.\mapping_\queries\prm{\query, \state} \equiv
\mapping^\prime\prm{\query, \theory{\state}}\]

Since \simulates[HMG+]{\system{T}}{\system{S}}, the HMG+ simulation contains the
mapping \(\pi : \queries^\system{S} \times \theory{\system{T}} \to \set{\true,
\false}\).

Thus, let \(\mapping^{\prime\prime}: \queries^\system{U} \times
\theory{\system{T}} \to \set{\true, \false}\) be constructed as follows. Use
\(\mapping^\prime: \queries^\system{U} \times \theory{\system{S}}\), and for
each query \query in \(\theory{\system{S}}\) that is needed by
\(\mapping^\prime\), consult \(\pi : \queries^\system{S} \times
\theory{\system{T}}\) to obtain a truth value in the current \system{T} state.
This forms a query decider that satisfies \propsym{QDt}.


Choose an arbitrary state \(\state^\system{U}_0 \in \states^\system{U}\) and
command \(\command^\system{U} \in \commands^\system{U}\), and let
\(\trans\prm{\state^\system{U}_0, \command^\system{U}} = \state^\system{U}_1\).
Let \(\state^\system{S}_0 \in \states^\system{S}\) such that
\(\state^\system{U}_0 \correspond[q] \state^\system{S}_0\). Since
\simulates[\properties]{\system{S}}{\system{U}},
\[\exists \state^\system{S}_1 \in
\states^\system{S}.(terminal\prm{\state^\system{S}_0,
\mapping_\commands\prm{\command^\system{U}, \state^\system{S}_0}} =
\state^\system{S}_1 \land \state^\system{U}_1 \correspond[q]
\state^\system{S}_1)\]

Let \(\state^\system{T}_0 \in \states^\system{T}\) such that
\(\state^\system{S}_0 \correspond[q] \state^\system{T}_0\). Since
\simulates[HMG+]{\system{T}}{\system{S}}, we can map each command of the
sequence \(\mapping_\commands\prm{\command^\system{U}, \state^\system{S}_0}\) to
a sequence of \system{T} commands using the HMG+ simulation. Concatenating this
sequence of sequences to a single sequence \(\commands^\system{T}_0\), and using
HMG+ correctness:
\[\exists \state^\system{T}_1 \in
\states^\system{T}.(terminal\prm{\state^\system{T}_0, \commands^\system{T}_0} =
\state^\system{T}_1 \land \state^\system{S}_1 \correspond[q]
\state^\system{T}_1)\]
Define \(\mapping_\commands: \commands^\system{U} \times \states^\system{T} \to
\prm{\commands^\system{T}}^*\) such that it returns \(\commands^\system{T}_0\)
for \(\state^\system{T}_0, \command^\system{U}\).

Then, given \(\state^\system{U}_0, \state^\system{U}_1 \in \states^\system{U},
\state^\system{T}_0 \in \states^\system{T}, \command^\system{U} \in
\commands^\system{U}\) such that \(\trans\prm{\state^\system{U}_0,
\command^\system{U}} = \state^\system{U}_1\), and \(\state^\system{U}_0
\correspond[q] \state^\system{T}_0\),
\[\exists \state^\system{T}_1 \in
\states^\system{T}.(terminal\prm{\state^\system{T}_0,
\mapping_\commands\prm{\command, \state^\system{T}_0}} = \state^\system{T}_1
\land \state^\system{S}_1 \correspond[q] \state^\system{T}_1)\]

Hence, \simulates[\propertyset]{\system{T}}{\system{U}}.
\end{proof}

\begin{lemma} \label{thm:hmg:if}
Given access control systems \system{S} and \system{T} and simulation properties
\(\properties = \propertyset\),
\(\reduction[\properties]{\system{S}}{\system{T}} \implies
\simulates[HMG+]{\system{T}}{\system{S}}\).
That is, if \system{T} is at least as expressive as \system{S} with respect to
properties \properties, then \system{T} admits an HMG+ simulation of \system{S}.
\end{lemma}

\begin{proof}
Let \system{S} and \system{T} be arbitrary access control systems such that
\reduction[\properties]{\system{S}}{\system{T}}. Since
\reduction[\properties]{\system{S}}{\system{T}}, for any access control system
\system{U}, if \simulates[\properties]{\system{S}}{\system{U}}, then
\simulates[\properties]{\system{T}}{\system{U}}.

Since \system{S} can trivially simulate itself,
\simulates[\properties]{\system{S}}{\system{S}}, and thus
\simulates[\properties]{\system{T}}{\system{S}}.

By \propsym{QDt},
\[\exists \mapping^\prime: \queries^\system{S} \times \theory{\system{T}} \to
\set{\true, \false}.\mapping_\queries\prm{\query^\system{S}, \state^\system{T}}
\equiv \mapping^\prime\prm{\query^\system{S}, \theory{\state^\system{T}}}\]
Thus, \(\mapping^\prime\) satisfies the format of the HMG+ query mapping (i.e.,
\(\pi : \queries^\system{S} \times \theory{\system{T}} \to \set{\true,
\false}\)), and by \propsym{SCq}, the state mapping preserves the query mapping,
property (i) for the HMG+ correct simulation.

Let \(\state^\system{S}_0 \in \states^\system{S}, \command \in
\commands^\system{S}\) be an arbitrary state and command in \system{S}, and
\(\state^\system{T}_0 \in \states^\system{T}\) a state in \system{T} such that
\(\mapping_\states\prm{\state^\system{S}_0} = \state^\system{T}_0\). Then, if
\(\trans\prm{\state^\system{S}_0, \command^\system{S}} = \state^\system{S}_1\),
\[\exists \state^\system{T}_1.(terminal\prm{\state^\system{T}_0,
\mapping_\commands\prm{\command, \state^\system{T}_0}} = \state^\system{T}_1
\land \state^\system{S}_1 \correspond[q] \state^\system{T}_1)\]

Thus, the command mapping preserves the state mapping, property (ii) for the
HMG+ correct simulation.

These properties satisfy the definition for HMG+ correct simulation, and hence
\system{T} admits an HMG+ simulation of \system{S}
(\simulates[HMG+]{\system{T}}{\system{S}}).
\end{proof}

\begin{theorem} \label{thm:hmg}
\(\decomposes{\text{HMG+}}{\propertyset}\); that is, the HMG+ parameterized
expressiveness simulation (correctness only) decomposes to query correspondence,
theory-dependent query, and forward reachability.
\end{theorem}

\begin{proof}
By \cref{thm:hmg:onlyif}, if \simulates[HMG+]{\system{T}}{\system{S}}, then
\reduction[\properties]{\system{S}}{\system{T}}. By \cref{thm:hmg:if}, if
\reduction[\properties]{\system{S}}{\system{T}}, then
\simulates[HMG+]{\system{T}}{\system{S}}. Thus,
\reduction[\properties]{\system{S}}{\system{T}} if and only if
\simulates[HMG+]{\system{T}}{\system{S}}, and thus the HMG+ simulation
decomposes to \propertyset.
\end{proof}

%% file: proofs/hmgetc.tex
\subsection{AC-Preserving HMG+ Parameterized Expressiveness}

\renewcommand{\propertyset}{\ensuremath{\set{\propsym{SCq}, \propsym{QDt}, \propsym{R\reach}, \propsym{QPa}}}\xspace}

For the purposes of this proof, let the set of simulation properties
\(\properties = \propertyset\).
\begin{lemma} \label{thm:hmga:onlyif}
Given access control systems \system{S}, \system{T}, and \system{U},
\[\simulates[HMG+a]{\system{T}}{\system{S}} \land
\simulates[\properties]{\system{S}}{\system{U}} \implies
\simulates[\properties]{\system{T}}{\system{U}}\]
That is, if \system{T} admits an HMG+ simulation with AC-preservation of
\system{S}, and \system{S} admits a simulation of \system{U} with properties
\propertyset, then \system{T} admits a simulation of \system{U} with properties
\propertyset.
\end{lemma}

\begin{proof}
The proof of \cref{thm:hmg:onlyif} proves all properties besides \propsym{QPa}
with no change. Thus, we now show \propsym{QPa}.


Choose an arbitrary request \(\request \in \requests^\system{U}\) and states
\(\state^\system{S} \in \states^\system{S}, \state^\system{T} \in
\states^\system{T}\). Since \simulates[\properties]{\system{S}}{\system{U}},
\[\mapping_\queries\prm{\request, \state^\system{S}} \equiv \state^\system{S}
\entails \request\]

Since \simulates[HMG+a]{\system{T}}{\system{S}},
\[\forall \request^\system{S} \in \requests^\system{S}:
\mapping_\queries\prm{\request^\system{S}, \state^\system{T}} \equiv
\state^\system{T} \entails \request^\system{S}\]

Since \simulates[\properties]{\system{S}}{\system{U}}, by \propsym{SCq},
\(\request \in \requests^\system{S}\). Thus,
\[\mapping_\queries\prm{\request, \state^\system{T}} \equiv \state^\system{T}
\entails \request\]

Hence, \simulates[\propertyset]{\system{T}}{\system{U}}.
\end{proof}

\begin{lemma} \label{thm:hmga:if}
Given access control systems \system{S} and \system{T} and simulation properties
\(\properties = \propertyset\),
\(\reduction[\properties]{\system{S}}{\system{T}} \implies
\simulates[HMG+a]{\system{T}}{\system{S}}\).
That is, if \system{T} is at least as expressive as \system{S} with respect to
properties \properties, then \system{T} admits an HMG+ simulation with
AC-preservation of \system{S}.
\end{lemma}

\begin{proof}
Let \system{S} and \system{T} be arbitrary access control systems such that
\reduction[\properties]{\system{S}}{\system{T}}. Since
\reduction[\properties]{\system{S}}{\system{T}}, for any access control system
\system{U}, if \simulates[\properties]{\system{S}}{\system{U}}, then
\simulates[\properties]{\system{T}}{\system{U}}.

Since \system{S} can trivially simulate itself,
\simulates[\properties]{\system{S}}{\system{S}}, and thus
\simulates[\properties]{\system{T}}{\system{S}}.

The proof of \cref{thm:hmg:if} proves all properties besides AC-preservation. By
\propsym{QPa},
\[\forall \request^\system{S} \in \requests^\system{S}:
\mapping_\queries\prm{\request^\system{S}, \state^\system{T}} \equiv
\state^\system{T} \entails \request^\system{S}\]

This satisfies the definition of AC-preservation, and hence \system{T} admits an
HMG+ simulation with AC-preservation of \system{S}
(\simulates[HMG+a]{\system{T}}{\system{S}}).
\end{proof}

\begin{theorem} \label{thm:hmga}
\(\decomposes{\text{HMG+a}}{\propertyset}\); that is, HMG+ parameterized
expressiveness with AC-preservation decomposes to query correspondence,
theory-dependent query, forward reachability, and authorization preservation.
\end{theorem}

\begin{proof}
By \cref{thm:hmga:onlyif}, if \simulates[HMG+a]{\system{T}}{\system{S}}, then
\reduction[\properties]{\system{S}}{\system{T}}. By \cref{thm:hmga:if}, if
\reduction[\properties]{\system{S}}{\system{T}}, then
\simulates[HMG+a]{\system{T}}{\system{S}}. Thus,
\reduction[\properties]{\system{S}}{\system{T}} if and only if
\simulates[HMG+a]{\system{T}}{\system{S}}, and thus the HMG+ simulation with
AC-preservation decomposes to \propertyset.
\end{proof}

\subsection{Monotonic HMG+ Parameterized Expressiveness}

\renewcommand{\propertyset}{\ensuremath{\set{\propsym{SCq}, \propsym{QDt}, \propsym{R\reach}, \propsym{CTa}}}\xspace}

For the purposes of this proof, let the set of simulation properties
\(\properties = \propertyset\).
\begin{lemma} \label{thm:hmgs:onlyif}
Given access control systems \system{S}, \system{T}, and \system{U},
\[\simulates[HMG+s]{\system{T}}{\system{S}} \land
\simulates[\properties]{\system{S}}{\system{U}} \implies
\simulates[\properties]{\system{T}}{\system{U}}\]
That is, if \system{T} admits a monotonic HMG+ simulation of \system{S}, and
\system{S} admits a simulation of \system{U} with properties \propertyset, then
\system{T} admits a simulation of \system{U} with properties \propertyset.
\end{lemma}

\begin{proof}
The proof of \cref{thm:hmg:onlyif} proves all properties besides \propsym{CTa}
with no change. Thus, we now show \propsym{CTa}.


Choose an arbitrary command \(\command^\system{U} \in \commands^\system{U}\) and
state \(\state^\system{S}_0 \in \states^\system{S}\). Let:
\begin{itemize}

\item \(\tup{\command^\system{S}_1, \ldots, \command^\system{S}_n} =
\mapping_\commands\prm{\command^\system{U}, \state^\system{S}_0}\)

\item \(\state^\system{S}_i = terminal\prm{\state^\system{S}_0,
\command^\system{S}_1 \circ \cdots \circ \command^\system{S}_i}\)

\end{itemize}

Since \simulates[\properties]{\system{S}}{\system{U}}, by \propsym{CTa}, this
sequence of states \(\state^\system{S}_i\) is monotonic.

Furthermore, let:
\begin{itemize}

\item \(\state^\system{T}_{0,0} = \mapping_\states\prm{\state^\system{S}_0}\)

\item \(\tup{\command^\system{T}_{i,1}, \ldots, \command^\system{T}_{i,m}} =
\mapping_\commands\prm{\command^\system{S}_i, \state^\system{T}_{i-1,0}}\)

\item \(\state^\system{T}_{i,0} = terminal\prm{\state^\system{T}_{i-1,0},
\command^\system{T}_{i,1} \circ \cdots \circ \command^\system{T}_{i,m}}\)

\item \(\state^\system{T}_{i,j} = terminal\prm{\state^\system{T}_{i,0},
\command^\system{T}_{i+1,1} \circ \cdots \circ \command^\system{T}_{i+1,j}}\)

\end{itemize}
Put simply, \(\command^\system{U}\) is simulated in \system{T} by the following
sequence of commands:
\[\command^\system{T}_{1,1}, \ldots, \command^\system{T}_{1,m},
\command^\system{T}_{2,1}, \ldots, \command^\system{T}_{n,m}\]
thus passing through the following trace of states:
\[\state^\system{T}_{0,0}, \ldots, \state^\system{T}_{0,m-1},
\state^\system{T}_{1,0}, \ldots, \state^\system{T}_{n-1,m-1},
\state^\system{T}_{n,0}\]

Since \simulates[HMG+s]{\system{T}}{\system{S}}, each subsequence
\(\state^\system{T}_{i,0}, \ldots, \state^\system{T}_{i+1,0}\) is monotonic.

Since \simulates[\properties]{\system{S}}{\system{U}}, by \propsym{CTa} and
\propsym{SCq}, the full sequence must also be monotonic.

Hence, \simulates[\propertyset]{\system{T}}{\system{U}}.
\end{proof}

\begin{lemma} \label{thm:hmgs:if}
Given access control systems \system{S} and \system{T} and simulation properties
\(\properties = \propertyset\),
\(\reduction[\properties]{\system{S}}{\system{T}} \implies
\simulates[HMG+s]{\system{T}}{\system{S}}\).
That is, if \system{T} is at least as expressive as \system{S} with respect to
properties \properties, then \system{T} admits a monotonic HMG+ simulation of
\system{S}.
\end{lemma}

\begin{proof}
Let \system{S} and \system{T} be arbitrary access control systems such that
\reduction[\properties]{\system{S}}{\system{T}}. Since
\reduction[\properties]{\system{S}}{\system{T}}, for any access control system
\system{U}, if \simulates[\properties]{\system{S}}{\system{U}}, then
\simulates[\properties]{\system{T}}{\system{U}}.

Since \system{S} can trivially simulate itself,
\simulates[\properties]{\system{S}}{\system{S}}, and thus
\simulates[\properties]{\system{T}}{\system{S}}.

The proof of \cref{thm:hmg:if} proves all properties besides monotonicity. Since
\propsym{CTa} satisfies the definition of monotonicity, we thus have that
\system{T} admits a monotonic HMG+ simulation of \system{S}
(\simulates[HMG+s]{\system{T}}{\system{S}}).
\end{proof}

\begin{theorem} \label{thm:hmgs}
\(\decomposes{\text{HMG+s}}{\propertyset}\); that is, HMG+ parameterized
expressiveness with monotonicity decomposes to query correspondence,
theory-dependent query, forward reachability, and access monotonicity.
\end{theorem}

\begin{proof}
By \cref{thm:hmgs:onlyif}, if \simulates[HMG+s]{\system{T}}{\system{S}}, then
\reduction[\properties]{\system{S}}{\system{T}}. By \cref{thm:hmgs:if}, if
\reduction[\properties]{\system{S}}{\system{T}}, then
\simulates[HMG+s]{\system{T}}{\system{S}}. Thus,
\reduction[\properties]{\system{S}}{\system{T}} if and only if
\simulates[HMG+s]{\system{T}}{\system{S}}, and thus the monotonic HMG+
simulation decomposes to \propertyset.
\end{proof}

\subsection{Admin-Preserving HMG+ Parameterized Expressiveness}

\renewcommand{\propertyset}{\ensuremath{\set{\propsym{SCq}, \propsym{QDt}, \propsym{R\reach}, \propsym{CAa}}}\xspace}

For the purposes of this proof, let the set of simulation properties
\(\properties = \propertyset\).
\begin{lemma} \label{thm:hmgp:onlyif}
Given access control systems \system{S}, \system{T}, and \system{U},
\[\simulates[HMG+p]{\system{T}}{\system{S}} \land
\simulates[\properties]{\system{S}}{\system{U}} \implies
\simulates[\properties]{\system{T}}{\system{U}}\]
That is, if \system{T} admits an admin-preserving HMG+ simulation of \system{S},
and \system{S} admits a simulation of \system{U} with properties \propertyset,
then \system{T} admits a simulation of \system{U} with properties \propertyset.
\end{lemma}

\begin{proof}
The proof of \cref{thm:hmg:onlyif} proves all properties besides \propsym{CAa}
with no change. Thus, we now show \propsym{CAa}.


Consider arbitrary command \(\command^\system{U} \in \commands^\system{U}\) and
state \(\state^\system{S} \in \states^\system{S}\). Since
\simulates[\properties]{\system{S}}{\system{U}}, by \propsym{CAa},
\[\forall \command^\system{S} \in \mapping_\commands\prm{\command^\system{U},
\state^\system{S}}, \alpha\prm{\command^\system{S}} \in A \implies
\alpha\prm{\command^\system{U}} \in A\]

Consider state \(\state^\system{T} \in \states^\system{T}\). Since
\simulates[HMG+p]{\system{T}}{\system{S}}, by admin-preservation,
\[\forall \command^\system{T} \in \mapping_\commands\prm{\command^\system{S},
\state^\system{T}}, \alpha\prm{\command^\system{T}} \in A \implies
\alpha\prm{\command^\system{S}} \in A\]

Thus,
\[\forall \command^\system{T} \in \mapping_\commands\prm{\command^\system{U},
\state^\system{T}}, \alpha\prm{\command^\system{T}} \in A \implies
\alpha\prm{\command^\system{U}} \in A\]

Hence, \simulates[\propertyset]{\system{T}}{\system{U}}.
\end{proof}

\begin{lemma} \label{thm:hmgp:if}
Given access control systems \system{S} and \system{T} and simulation properties
\(\properties = \propertyset\),
\(\reduction[\properties]{\system{S}}{\system{T}} \implies
\simulates[HMG+p]{\system{T}}{\system{S}}\).
That is, if \system{T} is at least as expressive as \system{S} with respect to
properties \properties, then \system{T} admits an admin-preserving HMG+
simulation of \system{S}.
\end{lemma}

\begin{proof}
Let \system{S} and \system{T} be arbitrary access control systems such that
\reduction[\properties]{\system{S}}{\system{T}}. Since
\reduction[\properties]{\system{S}}{\system{T}}, for any access control system
\system{U}, if \simulates[\properties]{\system{S}}{\system{U}}, then
\simulates[\properties]{\system{T}}{\system{U}}.

Since \system{S} can trivially simulate itself,
\simulates[\properties]{\system{S}}{\system{S}}, and thus
\simulates[\properties]{\system{T}}{\system{S}}.

The proof of \cref{thm:hmg:if} proves all properties besides admin-preservation.
Since \propsym{CAa} satisfies the definition of admin-preservation, we thus have
that \system{T} admits an admin-preserving HMG+ simulation of \system{S}
(\simulates[HMG+p]{\system{T}}{\system{S}}).
\end{proof}

\begin{theorem} \label{thm:hmgp}
\(\decomposes{\text{HMG+p}}{\propertyset}\); that is, HMG+ parameterized
expressiveness with admin-preservation decomposes to query correspondence,
theory-dependent query, forward reachability, and administration preservation.
\end{theorem}

\begin{proof}
By \cref{thm:hmgp:onlyif}, if \simulates[HMG+p]{\system{T}}{\system{S}}, then
\reduction[\properties]{\system{S}}{\system{T}}. By \cref{thm:hmgp:if}, if
\reduction[\properties]{\system{S}}{\system{T}}, then
\simulates[HMG+p]{\system{T}}{\system{S}}. Thus,
\reduction[\properties]{\system{S}}{\system{T}} if and only if
\simulates[HMG+p]{\system{T}}{\system{S}}, and thus the admin-preserving HMG+
simulation decomposes to \propertyset.
\end{proof}

%% file: proofs/smg.tex
\subsection{SMG Simulation}

\renewcommand{\propertyset}{\ensuremath{\set{\propsym{SCa}, \propsym{R\reach}}}\xspace}

For the purposes of this proof, let the set of simulation properties
\(\properties = \propertyset\).
\begin{lemma} \label{thm:smg:onlyif}
Given access control systems \system{S}, \system{T}, and \system{U},
\[\simulates[SMG]{\system{T}}{\system{S}} \land
\simulates[\properties]{\system{S}}{\system{U}} \implies
\simulates[\properties]{\system{T}}{\system{U}}\]
That is, if \system{T} admits an SMG simulation of \system{S}, and \system{S}
admits a simulation of \system{U} with properties \propertyset, then \system{T}
admits a simulation of \system{U} with properties \propertyset.
\end{lemma}

\begin{proof}
Let \system{S}, \system{T}, and \system{U} be arbitrary access control systems
such that \simulates[SMG]{\system{T}}{\system{S}} and
\simulates[\properties]{\system{S}}{\system{U}}. To prove \cref{thm:smg:onlyif},
we must then show that \simulates[\properties]{\system{T}}{\system{U}}.


Choose an arbitrary state \(\state^\system{U}_0 \in \states^\system{U}\) and
command \(\command^\system{U} \in \commands^\system{U}\), and let
\(\trans\prm{\state^\system{U}_0, \command^\system{U}} = \state^\system{U}_1\).
Let \(\state^\system{S}_0 \in \states^\system{S}\) such that
\(\state^\system{U}_0 \correspond[a] \state^\system{S}_0\). Since
\simulates[\properties]{\system{S}}{\system{U}},
\[\exists \state^\system{S}_1 \in
\states^\system{S}.(terminal\prm{\state^\system{S}_0,
\mapping_\commands\prm{\command^\system{U}, \state^\system{S}_0}} =
\state^\system{S}_1 \land \state^\system{U}_1 \correspond[a]
\state^\system{S}_1)\]

Let \(\state^\system{T}_0 \in \states^\system{T}\) such that
\(\state^\system{S}_0 \correspond[a] \state^\system{T}_0\). Since
\simulates[SMG]{\system{T}}{\system{S}}, request \request can be granted in
\system{S} if and only if \(\mapping\prm{\request}\) can be granted in
\system{T}. More concretely,
\[\exists \state^\system{T}_1 \in \states^\system{T}.(\state^\system{T}_0
\Mapsto \state^\system{T}_1 \land \state^\system{S}_1 \correspond[a]
\state^\system{T}_1)\]
Thus, there exists a sequence of \system{T} commands \(\commands^\system{T}_0\)
such that \(terminal\prm{\state^\system{T}_0, \commands^\system{T}_0} =
\state^\system{T}_1\). Define \(\mapping_\commands: \commands^\system{U} \times
\states^\system{T} \to \prm{\commands^\system{T}}^*\) such that it returns
\(\commands^\system{T}_0\) for \(\state^\system{T}_0, \command^\system{U}\).
This is formed by concatenating a sequence of sequences of commands: for each
command \(\command^\system{S}_i\) that \system{S} needs to execute to simulate
\(\command^\system{U}\), concatenate the commands that \system{T} needs to
execute to simulate \(\command^\system{S}_i\).

Then, given \(\state^\system{U}_0, \state^\system{U}_1 \in \states^\system{U},
\state^\system{T}_0 \in \states^\system{T}, \command^\system{U} \in
\commands^\system{U}\) such that \(\trans\prm{\state^\system{U}_0,
\command^\system{U}} = \state^\system{U}_1\), and \(\state^\system{U}_0
\correspond[a] \state^\system{T}_0\),
\[\exists \state^\system{T}_1 \in
\states^\system{T}.(terminal\prm{\state^\system{T}_0,
\mapping_\commands\prm{\command, \state^\system{T}_0}} = \state^\system{T}_1
\land \state^\system{S}_1 \correspond[a] \state^\system{T}_1)\]

Hence, \simulates[\propertyset]{\system{T}}{\system{U}}.
\end{proof}

\begin{lemma} \label{thm:smg:if}
Given access control systems \system{S} and \system{T} and simulation properties
\(\properties = \propertyset\),
\(\reduction[\properties]{\system{S}}{\system{T}} \implies
\simulates[SMG]{\system{T}}{\system{S}}\).
That is, if \system{T} is at least as expressive as \system{S} with respect to
properties \properties, then \system{T} admits an SMG simulation of \system{S}.
\end{lemma}

\begin{proof}
Let \system{S} and \system{T} be arbitrary access control systems such that
\reduction[\properties]{\system{S}}{\system{T}}. Since
\reduction[\properties]{\system{S}}{\system{T}}, for any access control system
\system{U}, if \simulates[\properties]{\system{S}}{\system{U}}, then
\simulates[\properties]{\system{T}}{\system{U}}.

Since \system{S} can trivially simulate itself,
\simulates[\properties]{\system{S}}{\system{S}}, and thus
\simulates[\properties]{\system{T}}{\system{S}}.

By \propsym{SCa} and \propsym{QPa}, request \request can be granted in
\system{S} if and only if \(\mapping\prm{\request}\) can be granted in
\system{T}. This satisfies the definition of the SMG simulation, and hence
\system{T} admits an SMG simulation of \system{S}
(\simulates[SMG]{\system{T}}{\system{S}}).
\end{proof}

\begin{theorem} \label{thm:smg}
\(\decomposes{\text{SMG}}{\propertyset}\); that is, the SMG simulation
decomposes to authorization correspondence and forward reachability.
\end{theorem}

\begin{proof}
By \cref{thm:smg:onlyif}, if \simulates[SMG]{\system{T}}{\system{S}}, then
\reduction[\properties]{\system{S}}{\system{T}}. By \cref{thm:smg:if}, if
\reduction[\properties]{\system{S}}{\system{T}}, then
\simulates[SMG]{\system{T}}{\system{S}}. Thus,
\reduction[\properties]{\system{S}}{\system{T}} if and only if
\simulates[SMG]{\system{T}}{\system{S}}, and thus the SMG simulation decomposes
to \propertyset.
\end{proof}

%% file: proofs/ganta.tex
\subsection{Ganta Simulation}

\renewcommand{\propertyset}{\ensuremath{\set{\propsym{SCa}, \propsym{QPa}, \propsym{CTs}, \propsym{R\bireach}}}\xspace}

For the purposes of this proof, let the set of simulation properties
\(\properties = \propertyset\).
\begin{lemma} \label{thm:ganta:onlyif}
Given access control systems \system{S}, \system{T}, and \system{U},
\[\simulates[Ganta]{\system{T}}{\system{S}} \land
\simulates[\properties]{\system{S}}{\system{U}} \implies
\simulates[\properties]{\system{T}}{\system{U}}\]
That is, if \system{T} admits a Ganta simulation of \system{S}, and \system{S}
admits a simulation of \system{U} with properties \propertyset, then \system{T}
admits a simulation of \system{U} with properties \propertyset.
\end{lemma}

\begin{proof}
Let \system{S}, \system{T}, and \system{U} be arbitrary access control systems
such that \simulates[Ganta]{\system{T}}{\system{S}} and
\simulates[\properties]{\system{S}}{\system{U}}. To prove
\cref{thm:ganta:onlyif}, we must then show that
\simulates[\properties]{\system{T}}{\system{U}}.


Choose an arbitrary state \(\state^\system{U}_0 \in \states^\system{U}\) and
command \(\command^\system{U} \in \commands^\system{U}\), and let
\(\trans\prm{\state^\system{U}_0, \command^\system{U}} = \state^\system{U}_1\).
Let \(\state^\system{S}_0 \in \states^\system{S}\) such that
\(\state^\system{U}_0 \correspond[a] \state^\system{S}_0\). Since
\simulates[\properties]{\system{S}}{\system{U}},
\[\exists \state^\system{S}_1 \in
\states^\system{S}.(terminal\prm{\state^\system{S}_0,
\mapping_\commands\prm{\command^\system{U}, \state^\system{S}_0}} =
\state^\system{S}_1 \land \state^\system{U}_1 \correspond[a]
\state^\system{S}_1)\]

Let \(\state^\system{T}_0 \in \states^\system{T}\) such that
\(\state^\system{S}_0 \correspond[a] \state^\system{T}_0\). Since
\simulates[Ganta]{\system{T}}{\system{S}}, by Property~2 there must exist an
equivalent history to \(\state^\system{S}_0 \Mapsto \state^\system{S}_1\) in
\system{T} with an access-correspondent completion state. Thus,
\[\exists \state^\system{T}_1 \in \states^\system{T}.(\state^\system{T}_0
\Mapsto \state^\system{T}_1 \land \state^\system{S}_1 \correspond[a]
\state^\system{T}_1)\]
Thus, there exists a sequence of \system{T} commands \(\commands^\system{T}_0\)
such that \(terminal\prm{\state^\system{T}_0, \commands^\system{T}_0} =
\state^\system{T}_1\). Define \(\mapping_\commands: \commands^\system{U} \times
\states^\system{T} \to \prm{\commands^\system{T}}^*\) such that it returns
\(\commands^\system{T}_0\) for \(\state^\system{T}_0, \command^\system{U}\).
This is formed by concatenating a sequence of sequences of commands: for each
command \(\command^\system{S}_i\) that \system{S} needs to execute to simulate
\(\command^\system{U}\), concatenate the commands that \system{T} needs to
execute to simulate \(\command^\system{S}_i\).

Then, given \(\state^\system{U}_0, \state^\system{U}_1 \in \states^\system{U},
\state^\system{T}_0 \in \states^\system{T}, \command^\system{U} \in
\commands^\system{U}\) such that \(\trans\prm{\state^\system{U}_0,
\command^\system{U}} = \state^\system{U}_1\), and \(\state^\system{U}_0
\correspond[a] \state^\system{T}_0\),
\[\exists \state^\system{T}_1 \in
\states^\system{T}.(terminal\prm{\state^\system{T}_0,
\mapping_\commands\prm{\command, \state^\system{T}_0}} = \state^\system{T}_1
\land \state^\system{S}_1 \correspond[a] \state^\system{T}_1)\]

Hence, \simulates[\set{\propsym{SCa},
\propsym{R\reach}}]{\system{T}}{\system{U}}. Next we show \propsym{R\bireach}.


Choose some arbitrary states \(\state^\system{T}_0, \state^\system{T}_1 \in
\states^\system{T}\) such that \(\state^\system{T}_0 \mapsto
\state^\system{T}_1\). Let \(\state^\system{S}_0 \in \states^\system{S}\) such
that \(\state^\system{S}_0 \correspond[q] \state^\system{T}_0\). Since
\simulates[Ganta]{\system{T}}{\system{S}}, by Property~3 there must exist an
equivalent history to \(\state^\system{T}_0 \mapsto \state^\system{T}_1\) in
\system{S} with an access-correspondent completion state. Thus,
\[\exists \state^\system{S}_1 \in \states^\system{S}.(\state^\system{S}_0
\Mapsto \state^\system{S}_1 \land \state^\system{S}_1 \correspond[a]
\state^\system{T}_1)\]

Let \(\state^\system{U}_0 \in \states^\system{U}\) such that
\(\state^\system{U}_0 \correspond[a] \state^\system{S}_0\). Since
\simulates[\properties]{\system{S}}{\system{U}},
\[\exists \state^\system{U}_1 \in \states^\system{U}.(\state^\system{U}_0
\Mapsto \state^\system{U}_1 \land \state^\system{U}_1 \correspond[a]
\state^\system{S}_1)\]

Thus, given \(\state^\system{T}_0, \state^\system{T}_1 \in \states^\system{T},
\state^\system{U}_0 \in \states^\system{U}\) such that \(\state^\system{T}_0
\mapsto \state^\system{T}_1\) and \(\state^\system{U}_0 \correspond[a]
\state^\system{T}_0\),
\[\exists \state^\system{U}_1 \in \states^\system{U}.(\state^\system{U}_0
\Mapsto \state^\system{U}_1 \land \state^\system{U}_1 \correspond[a]
\state^\system{T}_1)\]

Hence, \simulates[\set{\propsym{SCa},
\propsym{R\bireach}}]{\system{T}}{\system{U}}. Next we show \propsym{QPa}.


Choose an arbitrary request \(\request \in \requests^\system{U}\) and states
\(\state^\system{S} \in \states^\system{S}, \state^\system{T} \in
\states^\system{T}\). Since \simulates[\properties]{\system{S}}{\system{U}},
\[\mapping_\queries\prm{\request, \state^\system{S}} \equiv \state^\system{S}
\entails \request\]

Since \simulates[Ganta]{\system{T}}{\system{S}}, by Property~6,
\[\forall \request^\system{S} \in \requests^\system{S}:
\mapping_\queries\prm{\request^\system{S}, \state^\system{T}} \equiv
\state^\system{T} \entails \request^\system{S}\]

Since \simulates[\properties]{\system{S}}{\system{U}}, by \propsym{SCa},
\(\request \in \requests^\system{S}\). Thus,
\[\mapping_\queries\prm{\request, \state^\system{T}} \equiv \state^\system{T}
\entails \request\]

Hence, \simulates[\set{\propsym{SCa}, \propsym{QPa},
\propsym{R\bireach}}]{\system{T}}{\system{U}}. Next we show \propsym{CTs}.


Choose an arbitrary command \(\command^\system{U} \in \commands^\system{U}\) and
state \(\state^\system{S}_0 \in \states^\system{S}\). Let:
\begin{itemize}

\item \(\tup{\command^\system{S}_1, \ldots, \command^\system{S}_n} =
\mapping_\commands\prm{\command^\system{U}, \state^\system{S}_0}\)

\item \(\state^\system{S}_i = terminal\prm{\state^\system{S}_0,
\command^\system{S}_1 \circ \cdots \circ \command^\system{S}_i}\)

\end{itemize}

Since \simulates[\properties]{\system{S}}{\system{U}}, by \propsym{CTs},
\[Allowed\prm{\state^\system{S}_i} \subseteq Allowed\prm{\state^\system{S}_0}
\lor Allowed\prm{\state^\system{S}_i} \subseteq
Allowed\prm{\state^\system{S}_n}\]
for any \(\state^\system{S}_i\).

Furthermore, let:
\begin{itemize}

\item \(\state^\system{T}_{0,0} = \mapping_\states\prm{\state^\system{S}_0}\)

\item \(\tup{\command^\system{T}_{i,1}, \ldots, \command^\system{T}_{i,m}} =
\mapping_\commands\prm{\command^\system{S}_i, \state^\system{T}_{i-1,0}}\)

\item \(\state^\system{T}_{i,0} = terminal\prm{\state^\system{T}_{i-1,0},
\command^\system{T}_{i,1} \circ \cdots \circ \command^\system{T}_{i,m}}\)

\item \(\state^\system{T}_{i,j} = terminal\prm{\state^\system{T}_{i,0},
\command^\system{T}_{i+1,1} \circ \cdots \circ \command^\system{T}_{i+1,j}}\)

\end{itemize}
Put simply, \(\command^\system{U}\) is simulated in \system{T} by the following
sequence of commands:
\[\command^\system{T}_{1,1}, \ldots, \command^\system{T}_{1,m},
\command^\system{T}_{2,1}, \ldots, \command^\system{T}_{n,m}\]
thus passing through the following trace of states:
\[\state^\system{T}_{0,0}, \ldots, \state^\system{T}_{0,m-1},
\state^\system{T}_{1,0}, \ldots, \state^\system{T}_{n-1,m-1},
\state^\system{T}_{n,0}\]

Consider an arbitrary state \(\state^\system{T}_{i,j}\) in this trace. Since
\simulates[Ganta]{\system{T}}{\system{S}}, by Property~6
\[Allowed\prm{\state^\system{T}_{i,j}} \subseteq
Allowed\prm{\state^\system{T}_{i,0}} \lor Allowed\prm{\state^\system{T}_{i,j}}
\subseteq Allowed\prm{\state^\system{T}_{i+1,0}}\]

Since \simulates[\properties]{\system{S}}{\system{U}}, by \propsym{CTs} and
\propsym{SCa},
\[Allowed\prm{\state^\system{T}_{i,0}} \subseteq
Allowed\prm{\state^\system{T}_{0,0}} \lor Allowed\prm{\state^\system{T}_{i,0}}
\subseteq Allowed\prm{\state^\system{T}_{n,0}}\]
as well as for \(\state^\system{T}_{i+1,0}\) in place of
\(\state^\system{T}_{i,0}\).

Thus,
\[Allowed\prm{\state^\system{T}_{i,j}} \subseteq
Allowed\prm{\state^\system{T}_{0,0}} \lor Allowed\prm{\state^\system{T}_{i,j}}
\subseteq Allowed\prm{\state^\system{T}_{n,0}}\]

Hence, \simulates[\propertyset]{\system{T}}{\system{U}}.
\end{proof}

\begin{lemma} \label{thm:ganta:if}
Given access control systems \system{S} and \system{T} and simulation properties
\(\properties = \propertyset\),
\(\reduction[\properties]{\system{S}}{\system{T}} \implies
\simulates[Ganta]{\system{T}}{\system{S}}\).
That is, if \system{T} is at least as expressive as \system{S} with respect to
properties \properties, then \system{T} admits a Ganta simulation of \system{S}.
\end{lemma}

\begin{proof}
Let \system{S} and \system{T} be arbitrary access control systems such that
\reduction[\properties]{\system{S}}{\system{T}}. Since
\reduction[\properties]{\system{S}}{\system{T}}, for any access control system
\system{U}, if \simulates[\properties]{\system{S}}{\system{U}}, then
\simulates[\properties]{\system{T}}{\system{U}}.

Since \system{S} can trivially simulate itself,
\simulates[\properties]{\system{S}}{\system{S}}, and thus
\simulates[\properties]{\system{T}}{\system{S}}.

By \propsym{SCa}, \propsym{QPa}, and the definition of \(\mapping_\commands\),
we have a Ganta \emph{scheme mapping} from \system{S} to \system{T}, satisfying
Ganta simulation Property~1.

By \propsym{R\reach} and the definition of \(\mapping_\commands\), all histories
of \system{S} have equivalent serial histories of \system{T}, satisfying
Property~2.

By \propsym{R\bireach} and the definition of \(\mapping_\commands\), all
incomplete histories of \system{T} can be completed serially, and all complete
histories of \system{T} have equivalent histories of \system{S}, satisfying
Properties~3--5.

Finally, by \propsym{QPa} and \propsym{CTs}, completion states are
access-correspondent, and intermediate states are non-contaminating, satisfying
Property~6.

These properties define the Ganta simulation, and hence \system{T} admits a
Ganta simulation of \system{S} (\simulates[Ganta]{\system{T}}{\system{S}}).
\end{proof}

\begin{theorem} \label{thm:ganta}
\(\decomposes{\text{Ganta}}{\propertyset}\); that is, the Ganta simulation
decomposes to access correspondence, authorization preservation,
anti-contamination, and bidirectional reachability.
\end{theorem}

\begin{proof}
By \cref{thm:ganta:onlyif}, if \simulates[Ganta]{\system{T}}{\system{S}}, then
\reduction[\properties]{\system{S}}{\system{T}}. By \cref{thm:ganta:if}, if
\reduction[\properties]{\system{S}}{\system{T}}, then
\simulates[Ganta]{\system{T}}{\system{S}}. Thus,
\reduction[\properties]{\system{S}}{\system{T}} if and only if
\simulates[Ganta]{\system{T}}{\system{S}}, and thus the Ganta simulation
decomposes to \propertyset.
\end{proof}

%% file: proofs/cdmw.tex
\subsection{CDM Weak Simulation}

\renewcommand{\propertyset}{\ensuremath{\set{\propsym{SCa}, \propsym{QPa}, \propsym{CDi}, \propsym{R\reach}}}\xspace}

For the purposes of this proof, let the set of simulation properties
\(\properties = \propertyset\).
\begin{lemma} \label{thm:cdmw:onlyif}
Given access control systems \system{S}, \system{T}, and \system{U},
\[\simulates[CDMw]{\system{T}}{\system{S}} \land
\simulates[\properties]{\system{S}}{\system{U}} \implies
\simulates[\properties]{\system{T}}{\system{U}}\]
That is, if \system{T} admits a CDM weak simulation of \system{S}, and \system{S}
admits a simulation of \system{U} with properties \propertyset, then \system{T}
admits a simulation of \system{U} with properties \propertyset.
\end{lemma}

\begin{proof}
To prove this \lcnamecref{thm:cdmw:onlyif}, we let \system{S}, \system{T}, and
\system{U} be access control systems such that
\simulates[CDMw]{\system{T}}{\system{S}} and
\simulates[\properties]{\system{S}}{\system{U}} but are otherwise arbitrary, and
we show that \simulates[\properties]{\system{T}}{\system{U}}.


Choose an arbitrary state \(\state^\system{U}_0 \in \states^\system{U}\) and
command \(\command^\system{U} \in \commands^\system{U}\), and let
\(\trans\prm{\state^\system{U}_0, \command^\system{U}} = \state^\system{U}_1\).
Let \(\state^\system{S}_0 \in \states^\system{S}\) such that
\(\state^\system{U}_0 \correspond[a] \state^\system{S}_0\). Since
\simulates[\properties]{\system{S}}{\system{U}},
\[\exists \state^\system{S}_1 \in
\states^\system{S}.(terminal\prm{\state^\system{S}_0,
\mapping_\commands\prm{\command^\system{U}, \state^\system{S}_0}} =
\state^\system{S}_1 \land \state^\system{U}_1 \correspond[a]
\state^\system{S}_1)\]

Let \(\state^\system{T}_0 \in \states^\system{T}\) such that
\(\state^\system{S}_0 \correspond[a] \state^\system{T}_0\). Since
\simulates[CDMw]{\system{T}}{\system{S}},
\[\exists \state^\system{T}_1 \in \states^\system{T}.(\state^\system{T}_0
\Mapsto \state^\system{T}_1 \land \state^\system{S}_1 \correspond[a]
\state^\system{T}_1)\]
Thus, there exists a sequence of \system{T} commands \(\commands^\system{T}_0\)
such that \(terminal\prm{\state^\system{T}_0, \commands^\system{T}_0} =
\state^\system{T}_1\). Define \(\mapping_\commands: \commands^\system{U} \times
\states^\system{T} \to \prm{\commands^\system{T}}^*\) such that it returns
\(\commands^\system{T}_0\) for \(\state^\system{T}_0, \command^\system{U}\).

Then, given \(\state^\system{U}_0, \state^\system{U}_1 \in \states^\system{U},
\state^\system{T}_0 \in \states^\system{T}, \command^\system{U} \in
\commands^\system{U}\) such that \(\trans\prm{\state^\system{U}_0,
\command^\system{U}} = \state^\system{U}_1\), and \(\state^\system{U}_0
\correspond[a] \state^\system{T}_0\),
\[\exists \state^\system{T}_1 \in
\states^\system{T}.(terminal\prm{\state^\system{T}_0,
\mapping_\commands\prm{\command, \state^\system{T}_0}} = \state^\system{T}_1
\land \state^\system{S}_1 \correspond[a] \state^\system{T}_1)\]

Hence, \simulates[\set{\propsym{SCa},
\propsym{R\reach}}]{\system{T}}{\system{U}}. Next, we show \propsym{QPa}.


Choose an arbitrary request \(\request \in \requests^\system{U}\) and states
\(\state^\system{S} \in \states^\system{S}, \state^\system{T} \in
\states^\system{T}\). Since \simulates[\properties]{\system{S}}{\system{U}},
\[\mapping_\queries\prm{\request, \state^\system{S}} \equiv \state^\system{S}
\entails \request\]

Since \simulates[CDMw]{\system{T}}{\system{S}},
\[\forall \request^\system{S} \in \requests^\system{S}:
\mapping_\queries\prm{\request^\system{S}, \state^\system{T}} \equiv
\state^\system{T} \entails \request^\system{S}\]

Since \simulates[\properties]{\system{S}}{\system{U}}, by \propsym{SCa},
\(\request \in \requests^\system{S}\). Thus,
\[\mapping_\queries\prm{\request, \state^\system{T}} \equiv \state^\system{T}
\entails \request\]

Hence, \simulates[\set{\propsym{SCa}, \propsym{QPa},
\propsym{R\reach}}]{\system{T}}{\system{U}}. Next we show \propsym{CDi}.


Since \simulates[\properties]{\system{S}}{\system{U}}, by \propsym{CDi},
\(\exists \mapping^{CDi}: \commands^\system{U} \to
\prm{\commands^\system{S}}^*.(\mapping_\commands\prm{\command, \state} \equiv
\mapping^{CDi}\prm{\command})\)
Thus, \(\mapping_\commands\) maps \system{U} commands to \system{S} commands
without considering the state in which they will be executed.

Since \simulates[CDMw]{\system{T}}{\system{S}}, by weak model containment,
\system{S} commands are mapped to \system{T} commands without considering the
state in which they will be executed. Call this mapping \(\mapping^{CDM}\).

Thus, let \(\mapping^\prime : \commands^\system{U} \to
\prm{\commands^\system{T}}^* = \mapping^{CDM} \circ \mapping^{CDi}\), and say
\(\mapping_\commands\prm{\command^\system{U}, \state^\system{T}} \equiv
\mapping^\prime\prm{\command^\system{U}}\). This forms a command mapping that
satisfies \propsym{CDi}.

Hence, \simulates[\properties]{\system{T}}{\system{U}}.
\end{proof}

\begin{lemma} \label{thm:cdmw:if}
Given access control systems \system{S} and \system{T} and simulation properties
\(\properties = \propertyset\),
\(\reduction[\properties]{\system{S}}{\system{T}} \implies
\simulates[CDMw]{\system{T}}{\system{S}}\).
That is, if \system{T} is at least as expressive as \system{S} with respect to
properties \properties, then \system{T} admits a CDM weak simulation of
\system{S}.
\end{lemma}

\begin{proof}
To prove this \lcnamecref{thm:cdmw:if}, we let \system{S} and \system{T} be
arbitrary access control systems such that
\reduction[\properties]{\system{S}}{\system{T}}, and we show that
\simulates[CDMw]{\system{T}}{\system{S}}.

Since \reduction[\properties]{\system{S}}{\system{T}}, for any access control
system \system{U}, if \simulates[\properties]{\system{S}}{\system{U}}, then
\simulates[\properties]{\system{T}}{\system{U}}.

Since \system{S} can trivially simulate itself,
\simulates[\properties]{\system{S}}{\system{S}}, and thus
\simulates[\properties]{\system{T}}{\system{S}}.

Thus, given \(\state^\system{S}_0, \state^\system{S}_1 \in \states^\system{S},
\state^\system{T}_0 \in \states^\system{T}\), by \propsym{SCa} and
\propsym{R\reach}, if \(\state^\system{S}_0 \correspond[a] \state^\system{T}_0\)
and \(\state^\system{S}_0 \mapsto \state^\system{S}_1\), then
\[\exists \state^\system{T}_1.(\state^\system{T}_0 \Mapsto \state^\system{T}_1
\land \state^\system{S}_1 \correspond[a] \state^\system{T}_1)\]

By \propsym{CDi},
\[\exists \mapping^{CDi}: \commands^\system{S} \to
\prm{\commands^\system{T}}^*.(\mapping_\commands\prm{\command, \state} \equiv
\mapping^{CDi}\prm{\command})\]

This relation is thus a weak access-containment relation, which satisfies the
definition for a CDM weak simulation, and hence \system{T} admits a CDM weak
simulation of \system{S} (\simulates[CDMw]{\system{T}}{\system{S}}).
\end{proof}

\begin{theorem} \label{thm:cdmw}
\(\decomposes{\text{CDMw}}{\propertyset}\); that is, the CDM weak simulation
decomposes to authorization correspondence, authorization preservation,
independent command mapping, and forward reachability.
\end{theorem}

\begin{proof}
By \cref{thm:cdmw:onlyif}, if \simulates[CDMw]{\system{T}}{\system{S}}, then
\reduction[\properties]{\system{S}}{\system{T}}. By \cref{thm:cdmw:if}, if
\reduction[\properties]{\system{S}}{\system{T}}, then
\simulates[CDMw]{\system{T}}{\system{S}}. Thus,
\reduction[\properties]{\system{S}}{\system{T}} if and only if
\simulates[CDMw]{\system{T}}{\system{S}}, and thus the CDM weak simulation
decomposes to \propertyset.
\end{proof}

%% file: proofs/cdms.tex
\subsection{CDM Strong Simulation}

\renewcommand{\propertyset}{\ensuremath{\set{\propsym{SCa}, \propsym{QPa}, \propsym{CDi}, \propsym{CS1}, \propsym{R\reach}}}\xspace}

For the purposes of this proof, let the set of simulation properties
\(\properties = \propertyset\).
\begin{lemma} \label{thm:cdms:onlyif}
Given access control systems \system{S}, \system{T}, and \system{U},
\[\simulates[CDMs]{\system{T}}{\system{S}} \land
\simulates[\properties]{\system{S}}{\system{U}} \implies
\simulates[\properties]{\system{T}}{\system{U}}\]
That is, if \system{T} admits a CDM strong simulation of \system{S}, and
\system{S} admits a simulation of \system{U} with properties \propertyset, then
\system{T} admits a simulation of \system{U} with properties \propertyset.
\end{lemma}

\begin{proof}
To prove this \lcnamecref{thm:cdms:onlyif}, we let \system{S}, \system{T}, and
\system{U} be access control systems such that
\simulates[CDMs]{\system{T}}{\system{S}} and
\simulates[\properties]{\system{S}}{\system{U}} but are otherwise arbitrary, and
we show that \simulates[\properties]{\system{T}}{\system{U}}.


Choose an arbitrary state \(\state^\system{U}_0 \in \states^\system{U}\) and
command \(\command^\system{U} \in \commands^\system{U}\), and let
\(\trans\prm{\state^\system{U}_0, \command^\system{U}} = \state^\system{U}_1\).
Let \(\state^\system{S}_0 \in \states^\system{S}\) such that
\(\state^\system{U}_0 \correspond[a] \state^\system{S}_0\). Since
\simulates[\properties]{\system{S}}{\system{U}},
\[\exists \state^\system{S}_1 \in
\states^\system{S}.(terminal\prm{\state^\system{S}_0,
\mapping_\commands\prm{\command^\system{U}, \state^\system{S}_0}} =
\state^\system{S}_1 \land \state^\system{U}_1 \correspond[a]
\state^\system{S}_1)\]

Let \(\state^\system{T}_0 \in \states^\system{T}\) such that
\(\state^\system{S}_0 \correspond[a] \state^\system{T}_0\). Since
\simulates[CDMs]{\system{T}}{\system{S}},
\[\exists \state^\system{T}_1 \in \states^\system{T}.(\state^\system{T}_0
\mapsto \state^\system{T}_1 \land \state^\system{S}_1 \correspond[a]
\state^\system{T}_1)\]
Thus, there exists a \system{T} command \(\command^\system{T}_0\) such that
\(\trans\prm{\state^\system{T}_0, \command^\system{T}_0} =
\state^\system{T}_1\). Define \(\mapping_\commands: \commands^\system{U} \times
\states^\system{T} \to \commands^\system{T}\) such that it returns
\(\command^\system{T}_0\) for \(\state^\system{T}_0, \command^\system{U}\).

Then, given \(\state^\system{U}_0, \state^\system{U}_1 \in \states^\system{U},
\state^\system{T}_0 \in \states^\system{T}, \command^\system{U} \in
\commands^\system{U}\) such that \(\trans\prm{\state^\system{U}_0,
\command^\system{U}} = \state^\system{U}_1\), and \(\state^\system{U}_0
\correspond[a] \state^\system{T}_0\),
\[\exists \state^\system{T}_1 \in
\states^\system{T}.(\trans\prm{\state^\system{T}_0,
\mapping_\commands\prm{\command, \state^\system{T}_0}} = \state^\system{T}_1
\land \state^\system{S}_1 \correspond[a] \state^\system{T}_1)\]

Hence, \simulates[\set{\propsym{SCa},
\propsym{R\reach}}]{\system{T}}{\system{U}}. Next, we show \propsym{QPa}.


Choose an arbitrary request \(\request \in \requests^\system{U}\) and states
\(\state^\system{S} \in \states^\system{S}, \state^\system{T} \in
\states^\system{T}\). Since \simulates[\properties]{\system{S}}{\system{U}},
\[\mapping_\queries\prm{\request, \state^\system{S}} \equiv \state^\system{S}
\entails \request\]

Since \simulates[CDMs]{\system{T}}{\system{S}},
\[\forall \request^\system{S} \in \requests^\system{S}:
\mapping_\queries\prm{\request^\system{S}, \state^\system{T}} \equiv
\state^\system{T} \entails \request^\system{S}\]

Since \simulates[\properties]{\system{S}}{\system{U}}, by \propsym{SCa},
\(\request \in \requests^\system{S}\). Thus,
\[\mapping_\queries\prm{\request, \state^\system{T}} \equiv \state^\system{T}
\entails \request\]

Hence, \simulates[\set{\propsym{SCa}, \propsym{QPa},
\propsym{R\reach}}]{\system{T}}{\system{U}}. Next we show \propsym{CDi} and
\propsym{CS1}.


Since \simulates[\properties]{\system{S}}{\system{U}}, by \propsym{CDi} and
\propsym{CS1},
\(\exists \mapping^{CDi}: \commands^\system{U} \to
\commands^\system{S}.(\mapping_\commands\prm{\command, \state} \equiv
\mapping^{CDi}\prm{\command})\)
Thus, \(\mapping_\commands\) maps \system{U} commands to \system{S} commands
without considering the state in which they will be executed.

Since \simulates[CDMs]{\system{T}}{\system{S}}, by strong model containment,
\system{S} commands are mapped to single \system{T} commands without considering
the state in which they will be executed. Call this mapping \(\mapping^{CDM}\).

Thus, let \(\mapping^\prime : \commands^\system{U} \to \commands^\system{T} =
\mapping^{CDM} \circ \mapping^{CDi}\), and say
\(\mapping_\commands\prm{\command^\system{U}, \state^\system{T}} \equiv
\mapping^\prime\prm{\command^\system{U}}\). This forms a command mapping that
satisfies \propsym{CDi} and \propsym{CS1}.

Hence, \simulates[\properties]{\system{T}}{\system{U}}.
\end{proof}

\begin{lemma} \label{thm:cdms:if}
Given access control systems \system{S} and \system{T} and simulation properties
\(\properties = \propertyset\),
\(\reduction[\properties]{\system{S}}{\system{T}} \implies
\simulates[CDMs]{\system{T}}{\system{S}}\).
That is, if \system{T} is at least as expressive as \system{S} with respect to
properties \properties, then \system{T} admits a CDM strong simulation of
\system{S}.
\end{lemma}

\begin{proof}
To prove this \lcnamecref{thm:cdms:if}, we let \system{S} and \system{T} be
arbitrary access control systems such that
\reduction[\properties]{\system{S}}{\system{T}}, and we show that
\simulates[CDMs]{\system{T}}{\system{S}}.

Since \reduction[\properties]{\system{S}}{\system{T}}, for any access control
system \system{U}, if \simulates[\properties]{\system{S}}{\system{U}}, then
\simulates[\properties]{\system{T}}{\system{U}}.

Since \system{S} can trivially simulate itself,
\simulates[\properties]{\system{S}}{\system{S}}, and thus
\simulates[\properties]{\system{T}}{\system{S}}.

Thus, given \(\state^\system{S}_0, \state^\system{S}_1 \in \states^\system{S},
\state^\system{T}_0 \in \states^\system{T}\), by \propsym{SCa} and
\propsym{R\reach}, if \(\state^\system{S}_0 \correspond[a] \state^\system{T}_0\)
and \(\state^\system{S}_0 \mapsto \state^\system{S}_1\), then
\[\exists \state^\system{T}_1.(\state^\system{T}_0 \mapsto \state^\system{T}_1
\land \state^\system{S}_1 \correspond[a] \state^\system{T}_1)\]

By \propsym{CDi} and \propsym{CS1},
\[\exists \mapping^{CDi}: \commands^\system{S} \to
\commands^\system{T}.(\mapping_\commands\prm{\command, \state} \equiv
\mapping^{CDi}\prm{\command})\]

This relation is thus a strong access-containment relation, which satisfies the
definition for a CDM strong simulation, and hence \system{T} admits a CDM strong
simulation of \system{S} (\simulates[CDMs]{\system{T}}{\system{S}}).
\end{proof}

\begin{theorem} \label{thm:cdms}
\(\decomposes{\text{CDMs}}{\propertyset}\); that is, the CDM strong simulation
decomposes to authorization correspondence, authorization preservation,
independent command mapping, lock-step, and forward reachability.
\end{theorem}

\begin{proof}
By \cref{thm:cdms:onlyif}, if \simulates[CDMs]{\system{T}}{\system{S}}, then
\reduction[\properties]{\system{S}}{\system{T}}. By \cref{thm:cdms:if}, if
\reduction[\properties]{\system{S}}{\system{T}}, then
\simulates[CDMs]{\system{T}}{\system{S}}. Thus,
\reduction[\properties]{\system{S}}{\system{T}} if and only if
\simulates[CDMs]{\system{T}}{\system{S}}, and thus the CDM strong simulation
decomposes to \propertyset.
\end{proof}